\documentclass[12pt,preprint]{aastex}

\usepackage{amsmath}
\usepackage{graphicx}
\usepackage{amssymb}
\usepackage{amsmath}
\usepackage{mathrsfs}
\usepackage{natbib}
\usepackage[usenames]{color}

\shorttitle{Black Hole Accretion Efficiency}
\shortauthors{Beckwith et al.}

\begin{document}

\title{Estimating the Radiative Efficiency of Magnetized 
Accretion Disks Around Black Holes}

\author{Kris Beckwith and John F. Hawley}
\affil{Astronomy Department\\
University of Virginia\\ 
P.O. Box 400325\\
Charlottesville, VA 22904-4325}

\email{krb3u@virginia.edu; jh8h@virginia.edu}

\and
\author{Julian H. Krolik}
\affil{Department of Physics and Astronomy\\
Johns Hopkins University\\
Baltimore, MD 21218}

\email{jhk@pha.jhu.edu}

\begin{abstract}

Simulations of black hole accretion have shown that magnetic stresses
are present near and inside the innermost stable circular orbit
(ISCO).  This finding suggests that such flows may be more luminous than
predicted by the standard accretion disk model.  Here we apply a
prescription for heat dissipation within the simulated accretion flows
to estimate their implied radiative efficiency.  We assume that
dissipation is proportional to the current density squared, and find that
the resulting
azimuthally-averaged and shell-integrated radial profile is well-matched
to the radial heat dissipation profile of the standard disk model for the
region outside the ISCO, particularly when it is adjusted to account for
additional stress at the ISCO.  In contrast to the standard model, however,
the dissipation profile derived from the current density continues
past the ISCO and through the plunging region.  The total predicted
dissipation rate is between $\simeq 30\%$ and $\simeq 100\%$ greater
than that predicted by the standard model, depending on the black hole
spin.  Most of the additional dissipation takes place just outside 
the ISCO.  To predict luminosities, we assume instantaneous
radiation and zero optical depth, but allow for photon capture.
The net radiative efficiency seen by a distant observer is increased
relative to the standard model by $\simeq 25\%$--$80\%$, with the
largest fractional increase for intermediate black hole spins because
the increase in dissipation from enhanced stress that occurs for rapid
spin is partially offset by the increased likelihood
that the additional photons will be captured by the hole.

\end{abstract}


\section{Introduction}

Accretion onto black holes is the most efficient known method of
liberating the rest-mass energy of matter and it is generally accepted
that accretion through an orbiting disk is the engine that produces the
luminosity emerging from active galactic nuclei and Galactic X-ray
binary systems.  Increasingly sophisticated observations are now
providing detailed information that may soon lead to improved
understanding of the dynamics of this accreting gas, and, by inference,
the properties of the strong gravitational fields around black holes.

What has become the standard model for how mass arrives at black holes
and radiates light along the way was first presented in \cite{Novikov:1973}.
In this model, the flow is assumed to be time-steady, axisymmetric,
and geometrically thin, and follows circular Keplerian orbits at all
radii outside the innermost stable circular orbit (the ``ISCO", which
has radius $r_{ms}$).  Inter-ring
torques act to decrease slowly the specific angular momentum in the orbiting
gas, allowing accretion to occur.  At and inside the ISCO, these torques are
assumed to vanish, and whatever angular momentum and energy remain in
the gas is deposited in the central black hole; this boundary condition
determines the net flux of angular momentum throughout the time-steady disk.
Application of the principles of conservation of angular momentum and
energy then defines the radial profile of all vertically-integrated
properties of the accretion flow.  In particular, conservation
of energy demands that the difference between the work done in some
annulus by the torques and the net potential energy lost as matter flows
across the
annulus be deposited locally as heat.  The assumption of geometrical
thinness embedded in the standard model is equivalent to the assumption
that all the dissipated heat is radiated locally.  Because the
accretion inside the ISCO is assumed to be in free-fall, the integrated
luminosity per unit rest-mass accreted, i.e. the radiative efficiency of
accretion, is identically equal to the binding energy per unit rest-mass
at the ISCO.  This efficiency ranges from $\simeq 5.7\%$ for a
Schwarzschild black hole hole to as high as $\simeq 42\%$ for a
maximally spinning Kerr hole.

In the notation of \cite{Krolik:1999}, the radial dependence of
the dissipation rate is
\begin{equation}
  \label{eqn:NTdissprofile} Q = \frac{3 GM \dot{M}}{4 \pi r^{3}} R_{R}(r).
\end{equation} 
Here, $R_{R}(r)$ is a function encapsulating all of the
relativistic effects relevant for disk dynamics and the effect of the
(assumed) inner boundary condition. 
In the Newtonian limit for a Keplerian
potential, $R_{R}(r)$ takes the familiar form:  
\begin{equation}
  R_{R}(r) = 1 - \left( r_{m}/{r} \right)^{1/2},
\end{equation} 
where $r_{m}$ is the radius in the disk at which the stress is zero.
$Q$ is defined in the frame that comoves with the surfaces of the disk
(the ``disk frame").  Determining the form of this function as measured
by a distant observer requires a transformation into that observer's
reference frame (see \S\ref{sec:distrib}).

Although it does provide a context in which to understand many properties of a
wide variety of accretion systems, this standard model is now known to
have many limitations.  One of these limitations lies at the heart of
this paper: its assumption that stress, and therefore dissipation, end
at the ISCO.  This assumption is based on heuristic arguments,
framed in a purely hydrodynamic context
\cite[indeed][point out that strong magnetic fields
could undercut those arguments]{Page:1974}.   Now that we have strong reason to
believe that turbulent magnetic fields are responsible for accretion
\cite[]{Balbus:1998}, it is clear that dynamics near the ISCO must
be considered more carefully in order to determine the true radiative
efficiency \cite[]{Krolik:1999a,Gammie:1999}.   In a previous paper
\cite[]{Krolik:2005}, we showed that the magnetic stresses are,
in fact, quite strong at and inside the ISCO
\cite[in the language of][the ``stress edge" lies well inside the ISCO
and can even go right to the event horizon]{Krolik:2002}.  In this paper,
we examine
the consequences for energy output.  Similarly borrowing terminology
from \cite{Krolik:2002}, we seek to locate the ``radiation edge,"
the radius from within which little light emerges to reach the outside
world.

Our approach rests on a number of fully general relativistic three
dimensional MHD simulations of accretion onto black holes, reported in
a series of papers beginning with \cite{De-Villiers:2003b}.  These
simulations were focused primarily on dynamics and the torques
generated by Maxwell stresses within the accretion flow.  Because these
simulations were unable to track dissipation directly, we must begin by
developing a prescription to model the dissipation within the accretion
flow.  This prescription can then be used to predict the observed
luminosity.  To this end, we use the magnetic 4-current density as a
proxy for local heating in a fashion similar to that proposed by
\cite{Machida:2003} and \cite{Hirose:2004,Hirose:2005} to determine
where photons are generated, and then make use of a general
relativistic ray-tracing package described previously by
\cite{Beckwith:2004,Beckwith:2005} to follow the subsequent paths of
those photons.

The remainder of this paper is divided into four sections. In
\S\ref{sec:overview}, we review the features of the simulations on
which this work is based.  In \S\ref{sec:dissipation}, we outline how
the dissipation of gravitational potential energy within the flow may
be described by use of the magnetic 4-current density and compare this
quantity's radial profile to the dissipation profile expected from
the standard relativistic disk model.  In
\S\ref{sec:transform}, we show how these dissipation functions can
be transformed into radiation observed at large distance.  To do
so, we assume that all heat is radiated instantaneously and the
photons encounter no opacity.  This method is then applied in
\S\ref{sec:distrib} to generate photon
distribution functions and in \S\ref{sec:effective} to determine the observed
radiation profile and the effective radiative efficiency of the flow. Finally,
in \S\ref{sec:summ}, we summarize our results and point the way to
future work.

\section{Overview of Simulations}

\label{sec:overview}

The calculations presented here are based on the results of the
Keplerian Disk (KD) simulations, which have been analyzed in a series
of papers by \cite{De-Villiers:2003b}, \cite{Hirose:2004},
\cite{De-Villiers:2005} and \cite{Krolik:2005}. For
purposes of clarity, we present a brief summary of key aspects of these
simulations.  The equations of ideal non-radiative MHD are solved in
the static Kerr metric of a rotating black hole using Boyer-Lindquist
coordinates expressed in gravitational units $(G = M = c = 1)$ with
line element $ds^{2} = g_{tt} dt^{2} + 2g_{t \phi} dt d\phi + g_{rr}
dr^{2} + g_{\theta \theta} d \theta^{2} + g_{\phi \phi} d\phi^{2}$ and
signature $(-,+,+,+)$.  The determinant of the $4$-metric is $\alpha
\sqrt{\gamma}$, where $\alpha = (-g^{tt})^{-1/2}$ is the lapse function
and $\gamma$ is the determinant of the spatial $3$-metric.  The finite
difference algorithm used is described by \cite{De-Villiers:2003}.

The initial conditions for the KD simulations consist of an isolated
hydrostatic gas torus orbiting near the black hole, with the pressure
maximum located at $r \approx 25 M$ and a (slightly) sub-Keplerian
distribution of angular momentum throughout.  The initial magnetic
field is weak and purely poloidal, and follows isodensity surfaces
within the torus.  The MRI, acting on this initial field, leads to
large-amplitude MHD turbulence that drives the subsequent evolution of
this torus.  By the end of the simulation a quasi-steady state
accretion disk extends from the hole out to $r\sim 20M$.  Beyond this
radius the net mass motion shifts to outward flow as it absorbs angular
momentum from the inner disk.  In this paper, therefore, we focus our
attention on the region inside $20M$.

Here we will make use of the data from the four high resolution
simulations presented by \cite{De-Villiers:2003b}, which are designated
KD0, KDI, KDP and KDE and correspond to black hole spins of
$a=0.0,0.5,0.9,0.998$ respectively.  Each simulation used $192 \times
192 \times 64$ $(r,\theta,\phi)$ grid zones.  The radial grid extends
from an inner boundary located at $r_{in} = 2.05, 1.90, 1.45, 1.175 M$
for KD0, KDI, KDP, KDE (i.e. just outside the black hole event horizon
in each case), to the outer boundary located at $r_{out} = 120 M$ in
all cases.  The radial grid is graded according to a hyperbolic cosine
function in order to concentrate grid zones close to the inner
boundary.  An outflow condition is applied at both the inner and outer
radial boundary.  The $\theta$-grid spans the range $0.045 \pi \le
\theta \le 0.955 \pi$ using an exponential distribution that
concentrates zones near the equator.  A reflecting boundary condition
is used along the conical cutout surrounding the coordinate axis.
Finally, the $\phi$-grid spans the quarter plane, $0 \le \phi \le \pi /
2$, with periodic boundary conditions in $\phi$.  The use of this
restricted angular domain significantly reduces the computational
requirements of the simulation \cite[for further discussion of the
effects of this restriction
see][]{De-Villiers:2003a}.  Each simulation was run
to time $8100M$, which corresponds to approximately $10$ orbits at the
initial pressure maximum.  For each simulation the time step $\Delta t$
is determined by the extremal light crossing time for a zone on the
spatial grid and remains constant for the entire simulation
\cite[]{De-Villiers:2003}.

The state of the relativistic fluid at each point in spacetime is
described by its density $\rho$, specific internal energy $\epsilon$,
$4$-velocity $U^{\mu}$, and isotropic pressure $P$. The relativistic
enthalpy is $h = 1 + \epsilon + P / \rho$. The pressure is related to
$\rho$ and $\epsilon$ via the equation of state for an ideal gas, $P =
\rho \epsilon ( \Gamma - 1)$, where $\Gamma = 5/3$. The magnetic field
is described by two sets of variables.  The first is the constrained
transport magnetic field $\mathcal{B}^{i} = [ijk] F_{jk}$, where
$[ijk]$ is the completely anti-symmetric symbol, and $F_{jk}$ are the
spatial components of the electromagnetic field strength tensor.  From
these are derived the magnetic field four-vector, $(4\pi)^{1/2} b^{\mu}
=  ^{*}F^{\mu \nu} U_{\nu}$, and the magnetic field scalar,
$||b^{2}|| = b^{\mu} b_{\mu}$. The electromagnetic
component of the stress-energy tensor is
$T^{\mu \nu}_{\mathrm{(EM)}} = \frac{1}{2}g^{\mu \nu}
||b||^{2} + U^{\mu} U^{\nu} ||b||^{2} - b^{\mu} b^{\nu}$.

In this paper we shall examine the spatial distribution of the squared
current 4-vector $||J^{\mu}||^2$ as a stand-in for dissipation in the
accretion flow.  The behavior of $||J^{\mu}||^2$ in the KD simulations
was examined previously by \cite{Hirose:2004}.  The definition of the
current is
\begin{equation}
  \label{eqn:2.2} J^{\mu} = \frac{1}{4 \pi} \nabla_{\nu} F^{\mu \nu}.
\end{equation} 
Here $\nabla_{\nu}$ is the covariant derivative and
$F^{\mu \nu}$ is the electromagnetic field strength tensor. The
covariant derivative can be simplified by use of the anti-symmetry of
$F^{\mu \nu}$, yielding:  \begin{equation}
  \label{eqn:2.3} J^{\mu} = \frac{1}{4 \pi \sqrt{-g}} \partial_{\nu}
  \left( \alpha \sqrt{\gamma} g^{\mu \lambda} g^{\nu \xi} F_{\lambda
  \xi} \right) \end{equation} The current density scalar is then
obtained simply from $||J||^{2} = J^{\mu} J_{\mu}$.

Evaluation of
$J^{\mu}$ requires the computation of time-derivatives.  Since this
information was not saved directly during the simulation itself, 
$J^{\mu}$ is obtained using data saved from three adjacent time steps.
Considerable effort is required to generate $J^{\mu}$ for an
individual time-step within a simulation, so we limit our analysis in this
initial investigation to a snapshot of the current density at
$t=8080M$, where all but the $a=0.998$ case have reached a quasi-steady state.

In addition to the accretion disk, in all of these simulations unbound
matter flows out in a cone aligned with the rotation axis.  We have
found it to be useful to distinguish this unbound material from the bound
accretion flow.  We designate as outflow any cell in which the specific
energy $-h U_{t} > 1$ (signaling that it is unbound) and $U^r > 0$.
In this paper we focus exclusively on the ``bound'' material.

\section{Dissipation \& Relation to Current Density}

\label{sec:dissipation}

Determining the observational properties of an accretion disk model
requires locating where it generates heat, that is, measuring---rather
than assuming---the function $Q(r)$.  In a real disk, torques arise
from the action of Maxwell
stresses (and to a lesser extent Reynolds stresses) in the MHD
turbulence generated by the MRI.  The energy thus extracted from the
orbital motion is initially deposited into the kinetic and magnetic
energy on the largest scales of the turbulence.  Subsequently, this
energy cascades down to a dissipation scale (either viscous or
resistive) where it is finally thermalized.  The simulations describe
well the first stages of this process, but can only mimic indirectly
the last step: grid-level effects intervene at lengthscales far
longer than the likely physical scale of dissipation.

In the present simulations $Q$ cannot therefore be directly
determined.  The simulations use a nonconservative internal energy
equation, and much of the turbulent kinetic and magnetic energy is lost
numerically on the grid scale, or partially captured through the use
of an artificial viscosity.  We therefore cannot make rigorous
arguments in support of any particular function of the dynamical data
as a tracer of the local dissipation rate.  At present we must instead
look at simulation quantities that have the potential to trace the
dissipation, making plausibility arguments that may lead us towards a
good choice.

In doing so we have a powerful discriminant in the form of
equation~(\ref{eqn:NTdissprofile}).  So long as there is little global
energy transport by waves, the classical azimuthally-averaged and
vertically-integrated relationship between stress and dissipation still
holds even for time-dependent disks \cite[]{Balbus:1999}.  Thus,
equation~(\ref{eqn:NTdissprofile}) should still apply in the main disk
body if the instantaneous accretion rate is known, modulo a correction
that accounts for any modification in the angular momentum flux due to
magnetic stresses at and inside the ISCO \cite[]{Agol:2000}.
What we seek is more general: a measure of the dissipation rate that
may be used in the plunging region as well as the disk body, and can be
used in a sense that is local with respect to both position and time.
To this end, we reinterpret equation~(\ref{eqn:NTdissprofile}) as a
constraint on any candidate prescription for dissipation: a successful
proxy for dissipation must agree with it after azimuthal-averaging and
vertical-integration in the region (the main disk body) where that
equation is valid.

Our chosen proxy for tracing dissipation within the simulations is the
current density, $||J||^{2}$.  A naive application of Ohm's law (Power
$\propto$ Current$^{2}$) partially motivates this choice.
\cite{Rosner:1978} suggested that (in the case of the solar corona),
regions of high current density are \textit{candidates} for regions of
high magnetic dissipation (and hence thermal heating), as high current
density \textit{may} trigger anomalous resistivity through mechanisms
such as ion-acoustic turbulence, although it should be emphasized that
no physical model (in the context of accretion disks) relating current
density and dissipation is presently known.  The connections between
current and dissipation have also been examined in local MHD shearing
box simulations.  \cite{Hirose:2005} tracked dissipation
explicitly and showed that dissipation within a gas pressure-dominated
local shearing box is well correlated with the current density, but
$\propto |\mathbf{J}|^{1.13}$, not $|\mathbf{J}|^{2}$. (Recall that the
shearing box simulations are non-relativistic, and $\mathbf{J} = \nabla
\times \mathbf{B} / 4 \pi$.) In light of these results we have
considered arbitrary powers of current, $||J||^{x}$, and adjusted $x$
to provide a close match to the radial dependence of the standard $Q$
in the disk body.  We find that the choice $||J||^{2}$ is not only
physically well-motivated, but also provides the best match to the
slope of $Q$.  It is entirely possible that prescriptions involving
quantities other than the current density could serve as well or better
as indicators of the dissipation rate.  We have explored several
candidates, chosen on the basis of their relation to energy flow
within the disk (the local magnetic work, $U^\phi T^r_{\phi (EM)}$, for example),
but none passed our plausibility tests as well as the squared
current-density.

To make the match with equation~\ref{eqn:NTdissprofile} quantitatively,
first recall that the dissipation
rate, $Q$, is defined \cite[in][]{Page:1974} as the energy release per
unit proper time per unit proper area as measured by an observer who
co-rotates with the disk (the ``fluid frame").  However, the radiative
efficiency of the accretion flow (i.e., the ratio of the integral of
the dissipation rate to the rest mass energy of the accreted matter) is
defined as seen by a distant observer, at rest in the global
Boyer-Lindquist coordinate frame.  Determining the radiative efficiency
of the flow therefore requires transforming the dissipation rate from
the fluid frame to the distant observer frame.  For this reason, we
compute the dissipation profiles in the distant observer frame, as this
choice of frames provides the reader with the most direct link to the
integrated radiative efficiency.

Our specific procedure is to assume that the dissipation is due to
processes that act like a uniform resistivity (note
that the perfect MHD approximation on which the simulations were based
assumes that there is perfect conductivity in the fluid frame).  In the
language of \cite{Page:1974}, the radiation contribution to the
vertical energy flux $T^z_t$ at the top surface of the disk (as seen in
the coordinate frame) may be written as $-U_t q^z$, where $q^z$ is the
energy flux in the fluid frame.  Because $q^z$ is evaluated at the top
surface of the disk, it is implicitly derived from an integral with respect to
height of the local dissipation rate through the volume of the disk.  We
make the ansatz that the analog to $q^z$ in our situation 
(in which the disk is not razor-thin) is proportional to the
squared current density integrated over $\theta$, i.e.,
\begin{equation}
  \label{eqn:qz}
q^z \propto \int d\theta \sqrt{g_{\theta\theta}} ||J||^2.
\end{equation}
With this identification, the dissipation profile as a function of
radius in the coordinate frame is
\begin{equation}
  \label{eqn:dissprofile}
   D ( r ) \propto  \frac{\int_{\mathrm{bound}}
  {-U_{t}(r,\theta,\phi) ||J(r,\theta,\phi)||^{2} \alpha
\sqrt{\gamma} 
d\theta d\phi}}
  {\sqrt{g_{rr} (r, \theta = \pi /2) g_{\phi \phi} (r, \theta=\pi /2)}
  \int{d\phi}}.
\end{equation}

When computing a radial dissipation profile we confine our attention
to the disk-like material within the global flow, hence the use of
the subscript ``bound" on the integral to denote that it includes only
contributions from bound material within the accretion flow.
This excludes the region identified as the ``funnel wall jet'' in
previous studies \cite[]{De-Villiers:2005}.
The analogous quantity for the standard model \cite[]{Thorne:1974} is:
\begin{equation}
  \label{eqn:qdissprof}
   D ( r ) \propto  \frac{\int_{\mathrm{bound}}
  {-U_{t} Q \alpha \sqrt{\gamma} d\phi}}
  {\sqrt{g_{rr} (r, \theta = \pi /2) g_{\phi \phi} (r, \theta=\pi /2)} 
  \int{d\phi}}.
\end{equation}
Note that the factor $\alpha \sqrt{\gamma}$ in this last integral
is determined from the $(2+1)$-d line element given in \cite{Page:1974}.

In this context, however, there is another standard of comparison for
our dissipation prescription that is even more directly relevant. \cite{Agol:2000}
 generalized the \cite{Page:1974} solution to
allow for non-zero stress at $r_{ms}$.  Because magnetized accretion
disks do have stresses in the marginally stable region \cite[]{Krolik:2005},
we also construct a $D(r)$ for their solution in direct analogy with
equation~(\ref{eqn:qdissprof}).  The additional stress at $r_{ms}$
can be parameterized in terms of the additional integrated dissipation
in the disk $\Delta\epsilon$.  We call this generalized solution the
``stressed standard model".

\begin{figure}
  \begin{center}
  \leavevmode
  \includegraphics[width=0.40\textwidth]{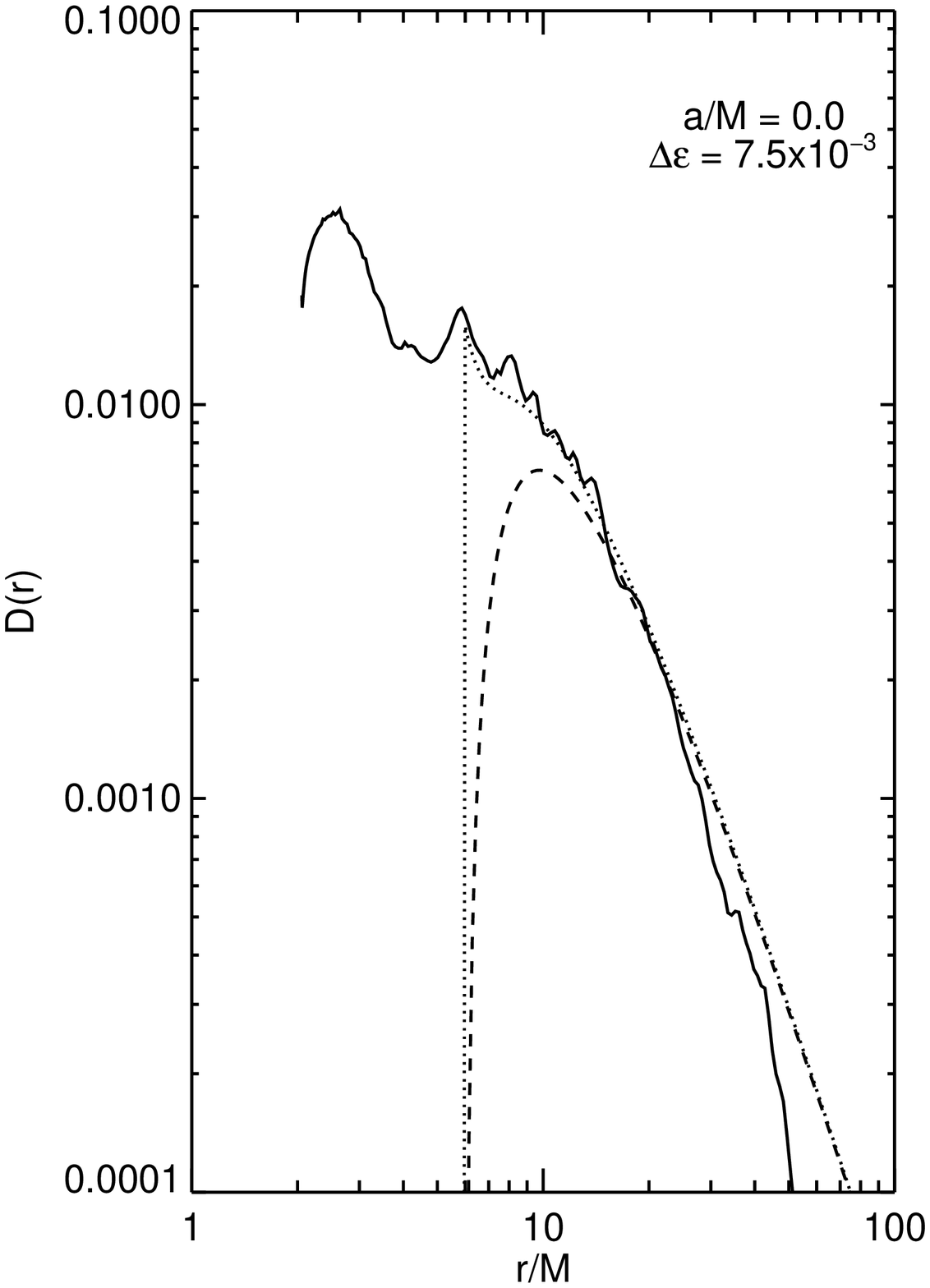}
  \includegraphics[width=0.40\textwidth]{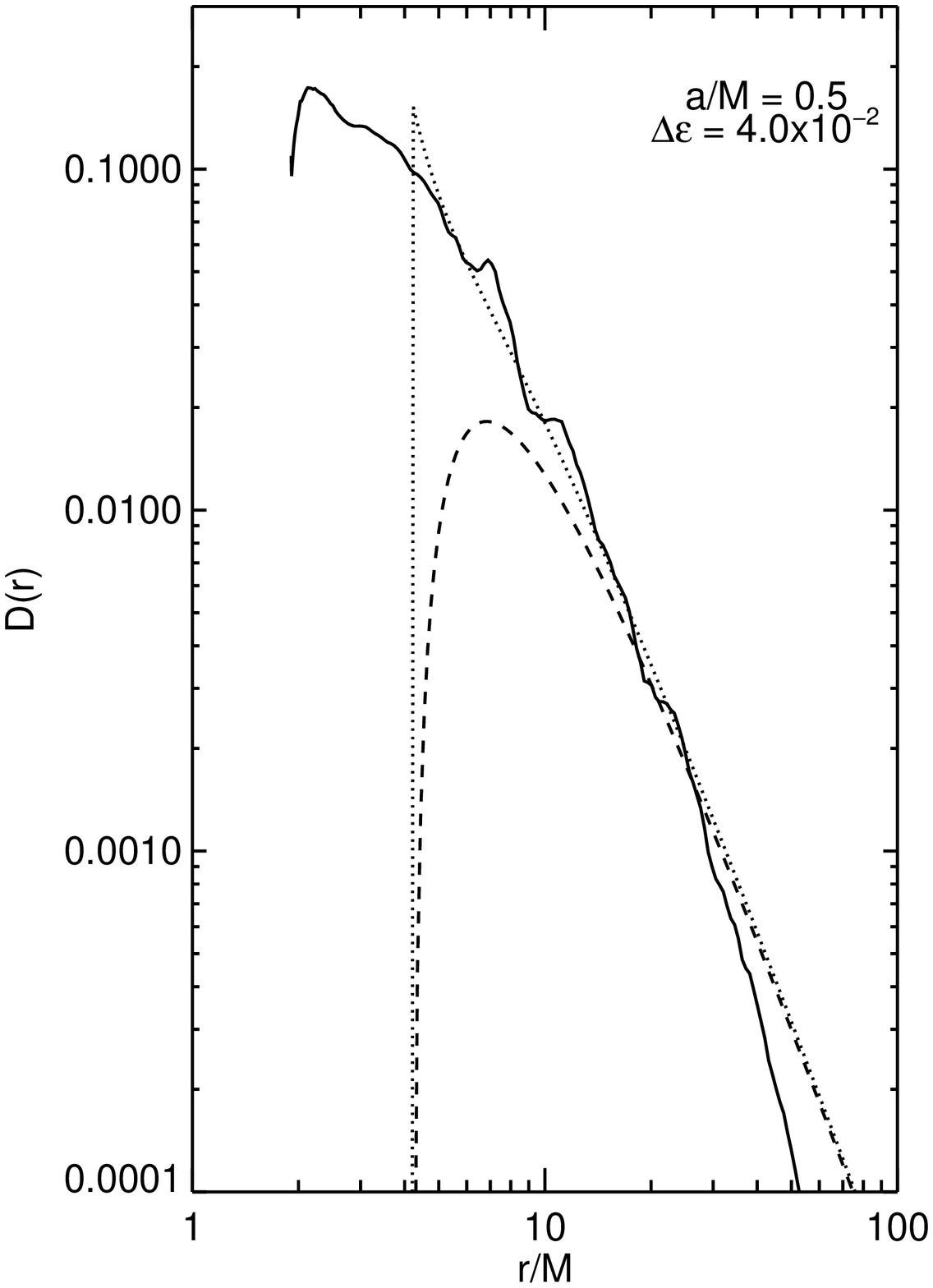}
  \includegraphics[width=0.40\textwidth]{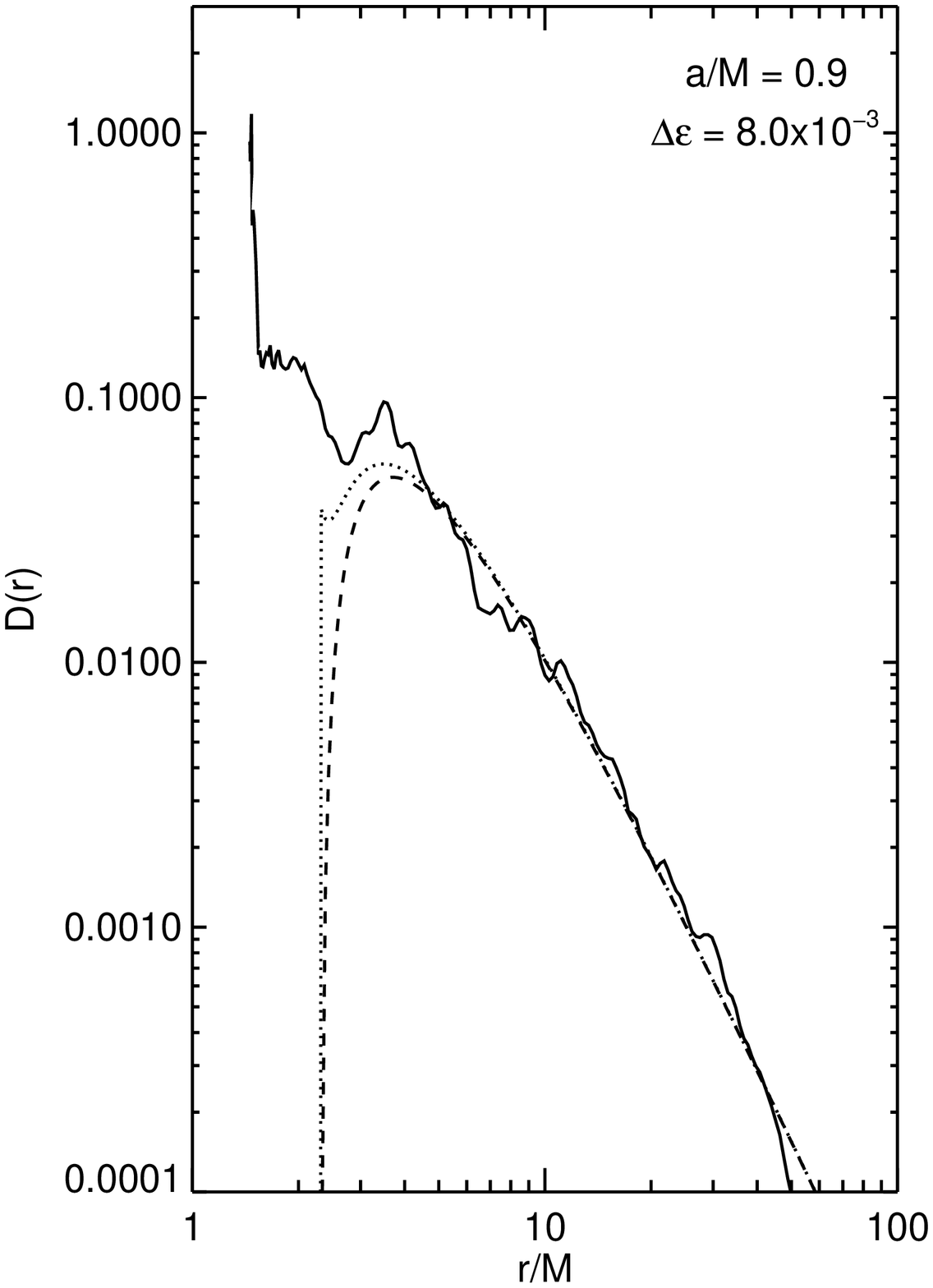}
  \includegraphics[width=0.40\textwidth]{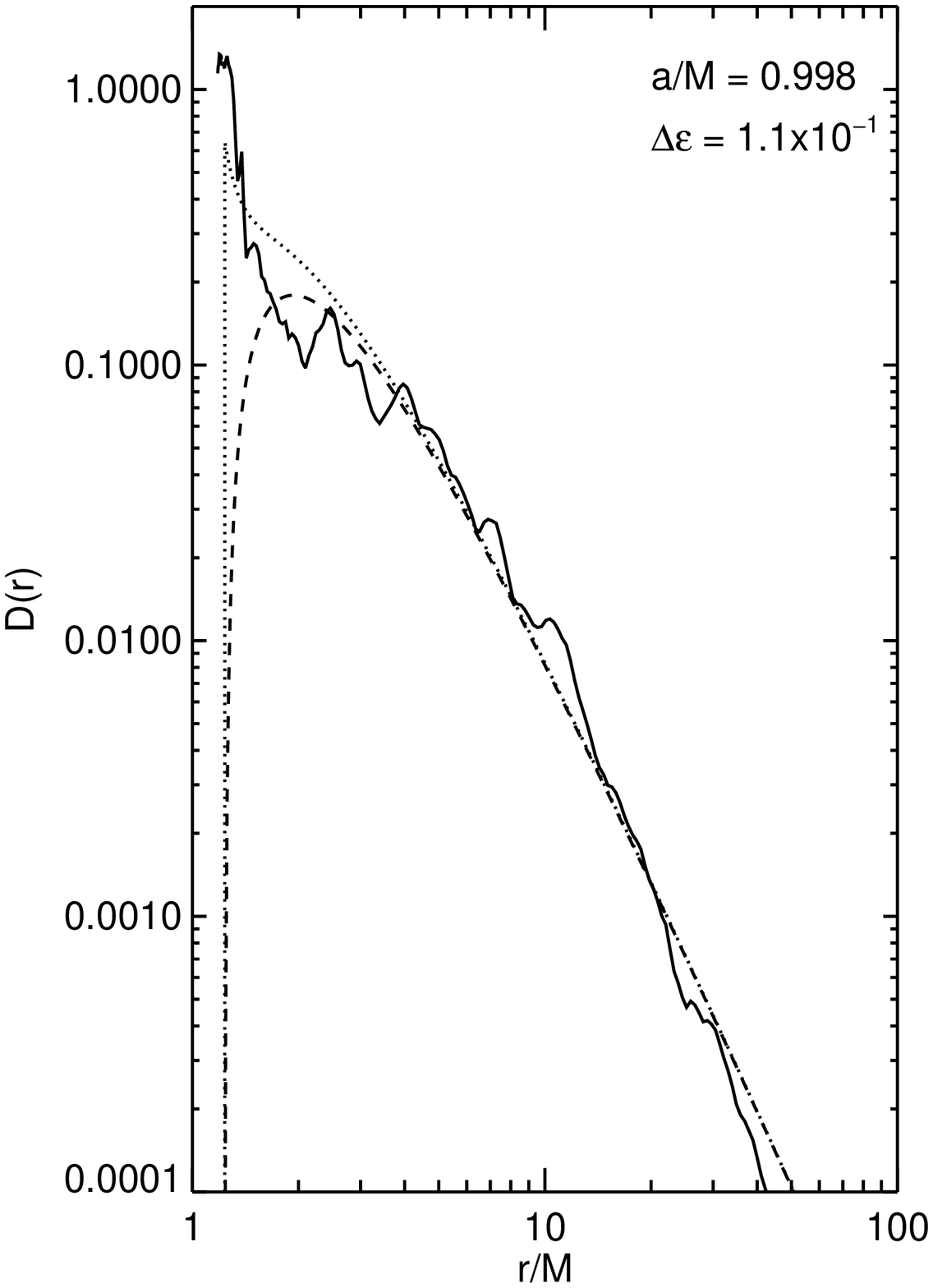}
  \end{center}
  \caption[]{Radial dissipation profile derived from the
  current density, $||J||^{2}$ (solid lines). 
  For comparison, we also show the dissipation function $Q$
  predicted by the standard relativistic disk model assuming the
  stress-free inner boundary condition at $r_{ms}$ (dashed lines), along with the stressed standard model proposed by \cite{Agol:2000} (dotted lines), where the choice of $\Delta \epsilon$ is given in the plot. The plots correspond to
  (a) KD0 ($a/M=0$); (b) KDI ($a/M=0.5$); (c) KDP ($a/M=0.9$) and (d) KDE
  ($a/M=0.998$). The radial profile of $||J||^{2}$ is normalized such that it matches $Q$ at $r=20$M.}
  \label{fig:candchecks} 
\end{figure}

In Figure~\ref{fig:candchecks}, we contrast three different versions
of $D(r)$: the standard model; the current
prescription with a coefficient chosen so that this version matches
the standard model at $r=20M$; and a ``stressed standard model" whose
additional stress is fixed by fitting to the current prescription
after its normalization to the standard model.  In the main disk
body, where there is little difference between the standard model
and the stressed standard model, we find that the current density
tracer does a good job of matching both curves.  As $r_{ms}$ is
approached from the outside, the $D(r)$ predicted by the current
density continues to
fit well the stressed standard model for all four black hole spins.
Most strikingly, the dissipation profile predicted by the current
varies smoothly across $r_{ms}$, in marked contrast to the predictions
of either the standard or the stressed standard disk model.  The
former goes to zero at $r_{ms}$ as a consequence of the
zero-stress inner boundary condition; the latter does the same because
it is not defined at smaller radii.  In addition, not only is $D(r)$
continuous across $r_{ms}$, it is also continuous through the
plunging region down to the inner boundary of the simulation.  This behavior
implies that, for a given black hole mass, spin, and accretion rate, the
accretion flow is more efficient in dissipating energy within the flow
than would be expected from the standard accretion disk model.
We must remind the reader, however, that although the three lower-spin
simulations had achieved a statistical steady-state by $t=8080M$
in the sense that their shell-integrated mass accretion rates varied
very little with radius, it is clear from the radial fluctuations in
these curves that a great many more samples would be necessary in order
to achieve a good estimate of the time-average shell-integrated
$||J||^2$.

Dissipation over and above the standard model prediction can occur
both inside and outside $r_{ms}$.
Data shown in Table \ref{tab:currentdiss} give some
sense of just how much additional dissipation there can be in both the plunging
region and the stably-orbiting disk.  To compute these numbers, we  define 
\begin{equation}
  \label{eqn:3.6}
L(r_{1},r_{2}) = \int^{r_{2}}_{r_{1}} 
\int^{\phi} D (r) \sqrt{g_{rr} g_{\phi \phi}} dr d\phi
\end{equation}
as the total energy dissipation rate within the region $r_{1} < r <
r_{2}$, where $D(r)$ is the dissipation function derived either from
the current or the magnetic work. For a given dissipation function,
we compute $L$ for two regions:
inside the plunging region, $r_{in} \le r < r_{ms}$, and within that part
of the stably-orbiting disk where accretion is taking place,
$r_{ms} \le r \le 20M$. In reality, dissipation will take place at radii
far in excess of $20M$, beyond the limit of the (simulated) stably-orbiting
disk. Integrating the standard model dissipation rate, $Q$ for
$r_{ms} \le r < 20M$ and $20M \le r < \infty$ reveals that, for $a/M = 0.0$,
more than \emph{half} of the total energy liberated from the accretion flow
originates outside $20M$.  A fair estimate of the relative increase in
the radiative efficiency created by the additional dissipation must therefore
include an estimate of dissipation outside $r = 20M$.
To accomplish this, we adopt a
simple prescription: for $r \ge 20M$, the simulated dissipation rates are
given by $Q$.

Table \ref{tab:currentdiss} shows the data for dissipation derived from the
current.  Depending on the spin parameter, the total dissipation rate is
between $30\%$ and $100\%$ greater than predicted by the standard model, with the maximum occurring for intermediate spin $a/M=0.5$.
Most of this additional dissipation takes place in the inner disk not
far outside the ISCO, with the plunging region contributing an interesting,
but distinctly minority, share (between $13\%$ and $27\%$ of the
total).  Thus, if the current density correctly gives the heating rate, the
standard model substantially underestimates its magnitude.

\begin{deluxetable}{lrcccccc}
\tablecolumns{5}
\tablewidth{0pc}
\tablecaption{Relative Dissipation Rates Derived from $||J||^{2}$}
\tablehead{\colhead{Model}          &
            \colhead{$a/M$}         &
            \colhead{$\frac{L(r_{in},r_{ms})}{L(r_{in},\infty)} $}         &
            \colhead{$\frac{L(r_{ms},\infty)}{L(r_{in},\infty)}$} &
            \colhead{$\frac{L(r_{in},\infty)}{L_{SM}(r_{ms},\infty)} $} &
}
\startdata
KD0 &    0.0 &      0.13 &      0.87 &       1.30 \\
KDI &      0.5 &      0.25 &      0.75 &       1.99 \\
KDP &      0.9 &      0.27 &      0.73 &       1.52  \\
KDE &      0.998 &      0.14 &      0.86 &       1.60 \\
\enddata
\label{tab:currentdiss}
\end{deluxetable}

So far we have discussed only the ratio of the predicted dissipation
to that of the standard model.  It is also useful to look
at their absolute levels (see Table~\ref{tab:absdiss}).  In this
table, $\epsilon_{SM}$ denotes the total dissipation predicted by
the classical Novikov-Thorne model, integrated from $r_{ms}$ to
$\infty$.  This quantity is what is generally called the radiative
efficiency.  The subscript $SSM$ denotes ``stressed standard model,''
by which we mean the standard model adjusted for supplemental
stress at $r_{ms}$ \cite[]{Agol:2000}, with the additional stress
chosen by fitting to the current-density curve.  The last
column is the integrated dissipation over the entire accretion
flow as predicted by the current-density estimator.  For the most part, this
new estimate of the total dissipation is considerably larger
than the standard model, raising its nominal radiative efficiency
$30\%$ (the non-rotating case), $100\%$ ($a/M = 0.5$), or
$50-60\%$ (the two high-spin cases).  Comparing $\epsilon_{SSM}$ to
$\epsilon_{||J||^2}$ shows that
most, but not all, of the additional dissipation takes place
outside $r_{ms}$.   In all cases, as we will show in detail
in the later sections of this paper, the integrated dissipation rate
is greater than the luminosity actually reaching infinity.

\begin{deluxetable}{lrccccc}
\tablecolumns{5}
\tablewidth{0pc}
\tablecaption{Radiative Efficiencies of Different Dissipation Models}
\tablehead{\colhead{Model}          &
            \colhead{$a/M$}         &
            \colhead{$\epsilon_{SM} (r_{ms},\infty)$}         &
            \colhead{$\epsilon_{SSM} (r_{ms},\infty)$}         &
            \colhead{$\epsilon_{||J||^{2}} (r_{in},\infty)$}         &
}
\startdata
KD0 &    0.0 &     0.06 &    0.07  &  0.08 \\
KDI &      0.5 &     0.09   &  0.13  &   0.17 \\
KDP &      0.9 &      0.16  &   0.17  &  0.25 \\
KDE &      0.998 &      0.34  &   0.45   &   0.54 \\
\enddata

\label{tab:absdiss}
\end{deluxetable}

\begin{figure}
  \begin{center}
  \leavevmode
  \includegraphics[width=\textwidth]{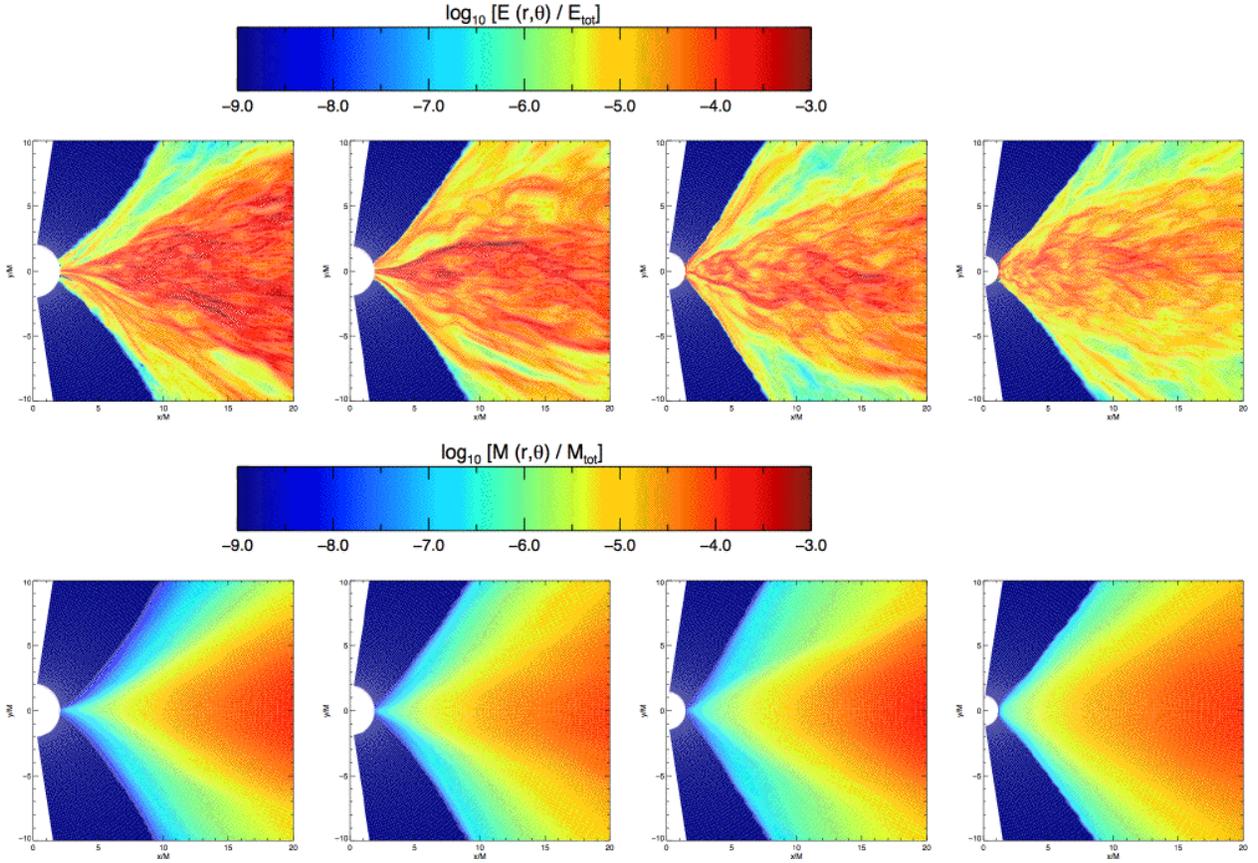}
  \end{center}
  \caption[Maps of the Heating Distribution for the KD
  Simulations]{Maps of the heating distribution 
  $E ( r, \theta )$ derived from $||J||^2$ along with the mass distribution $M(r,\theta)$ for $r < 20M$. 
  From left to right, the panels
  are maps for the KD0, KDI, KDP and KDE simulations, which
  correspond to black hole spins of $a = 0.0, 0.5, 0.9, 0.998$
  respectively. The top row of panels show $E ( r, \theta )$;
  the bottom row of panels show $M(r,\theta)$. In each case, the energy
  released is dominated by contributions within the main disk body,
  especially in the region around the equatorial plane.}
  \label{fig:heatingmaps} 
\end{figure}

One potential advantage of global numerical simulations over the
standard model is that more complex (and perhaps more realistic)
emission geometries are possible.  In the standard geometrically thin
disk model, all emission originates in the equatorial plane.  By
contrast, the disks in the KD simulations are somewhat geometrically
thick (typical ratio of matter scale-height to radius $\simeq 0.1$),
and so emission can occur away from the equatorial plane.  Using the
current density prescription for dissipation, we find that the
azimuthal average of the total energy released in a cell centered on
$r,\theta$ is given by:
\begin{equation}
  \label{eqn:2.6}
   E( r, \theta ) \propto
    \int^{\pi / 2}_{0} {-U_{t}( r, \theta, \phi ) ||J( r, \theta, \phi )||^{2} \alpha \sqrt{\gamma} \Delta t \Delta r 
    \theta d\phi} 
\end{equation}
The total energy released from a given simulation is defined as
$E_{\mathrm{tot}} = \sum_{\mathrm{bound}} {E (r,\theta)}$. Figure
\ref{fig:heatingmaps} shows a map of this heating distribution for each
of the KD simulations, where we have again removed contributions from
unbound material. For reference, we also show the spatial distribution
of total mass, given by:
\begin{equation}
   M( r, \theta ) \propto
    \int^{\pi / 2}_{0} {\rho( r, \theta, \phi ) \alpha \sqrt{\gamma} \Delta t \Delta r \Delta
    \theta d\phi} 
\end{equation}
The total mass within the disk body is similarly defined as
$M_{\mathrm{tot}} = \sum_{\mathrm{bound}} {M (r,\theta)}$. Examining
Figure \ref{fig:heatingmaps}, we see that the majority of energy
release is concentrated around the equatorial plane, where the bulk of
material in the disk is located.

It is of particular interest to study the division of dissipation
between high-density and low-density regions because the local density
is likely to be play an important role in determining the character of
the radiation ultimately produced.  Where the density is high,
thermalization is the most likely fate of the dissipated energy.
In such an environment, the absorptive opacity is typically high,
matter and radiation readily exchange energy, and the emergent
spectrum is near blackbody.  On the other hand, release of heat in
low-density regions may lead to temperatures so high that absorptive
opacity becomes minimal, and the characteristic emergent radiation
is hard X-rays produced by inverse Compton scattering.

Here we find that, if dissipation is correctly signalled by the
squared current density, it mostly takes place in the dense disk
body.  However, low-density regions receive a somewhat disproportionate
share, in the sense that the mean rate of heat release per unit mass
rises as the density falls.  Table~\ref{tab:coronadiss} makes this
point quantitatively.  For the purposes of this table, we define the
disk ``corona" as those regions of bound matter where the density is less
than $e^{-2}$ times the density on the midplane at that radius and
azimuthal angle.  As it shows, the fraction of heat deposited in
the corona defined in this manner falls in the range 15--$30\%$,
with a possible rising trend with increasing black hole
spin.  Moreover, the rate of dissipation per unit mass is between
$\simeq 2.5$ and 5 times greater than the mean for the disk body inside $20$M.

\begin{deluxetable}{lrcccccc}
\tablecolumns{5}
\tablewidth{0pc}
\tablecaption{Energy dissipation in coronal regions}
\tablehead{\colhead{Model}          &
            \colhead{$a/M$}         &
            \colhead{$F_{E} = E_{cor} / E_{tot}$} &
            \colhead{$F_{M} = M_{cor} / M_{tot}$} &
            \colhead{$F_{E} / F_{M}$} & }
\startdata
KD0 &       0.0 &      0.15 &     0.05 &       3.1 \\
KDI  &      0.5 &       0.17 &     0.07 &      2.6 \\
KDP &      0.9 &       0.29 &     0.06 &      5.1 \\
KDE &      0.998 &   0.21 &     0.04 &      4.8 \\
\enddata
\label{tab:coronadiss}
\end{deluxetable}

In this section we have
considered dissipative efficiency derived from 
a physically plausible dissipation process.  Despite the significant uncertainties, our results indicate
that dissipation in the marginally stable region can be significant,
and that the nominal radiative efficiency of accretion onto black
holes predicted by the standard model may be a substantial
underestimate.

\section{Transforming the Dissipation Profile}

\label{sec:transform}

If we wish to relate the proposed link between current density and
dissipation to the gross \emph{observed} energetics of magnetized
accretion flows, we must account for several
processes that lie between local heat dissipation and the detection of
light by distant observers. The dissipated heat must be converted to
photons that must make their way through any local opacity  before
arriving at infinity with their energies altered by both Doppler and
gravitational effects.  Some of the radiated photons can be captured
by the black hole, never reaching infinity at all.  In fact, even
in the standard model, photon capture makes the {\it effective} radiative
efficiency smaller than the nominal efficiency.

Here we take a greatly-simplified approach to the problem. We use
$||J||^{2}$ as a local tracer of dissipation and assume that heat is
instantly converted into photons that are emitted isotropically in the
rest frame of the fluid.  We further assume that the disk material is
optically thin, so that this  radiation escapes from the matter
instantaneously and is not advected with the flow. We follow the
photons' trajectories until they are captured by the hole or reach
infinity.  For those photons that reach infinity, we calculate the
photon transfer function (the ratio between the intensity as received
at infinity and as radiated in the fluid frame: see
eqn.~\ref{eqn:obsprofile}) and note how many times each path crosses
the equatorial plane.  Finally, for a given radius in the accretion
flow, we convolve the photon transfer function with $||J||^{2}$ and
integrate the resulting function over energy and observed
inclination.   This yields a solid angle- and frequency-integrated
observed luminosity for a given radius within the accretion flow.
Summing over radii, we obtain the observed luminosity.  Its ratio to
the rest-mass accretion rate is the effective radiative efficiency.

\begin{figure}
  \begin{center}
  \leavevmode
  \includegraphics[width=\textwidth]{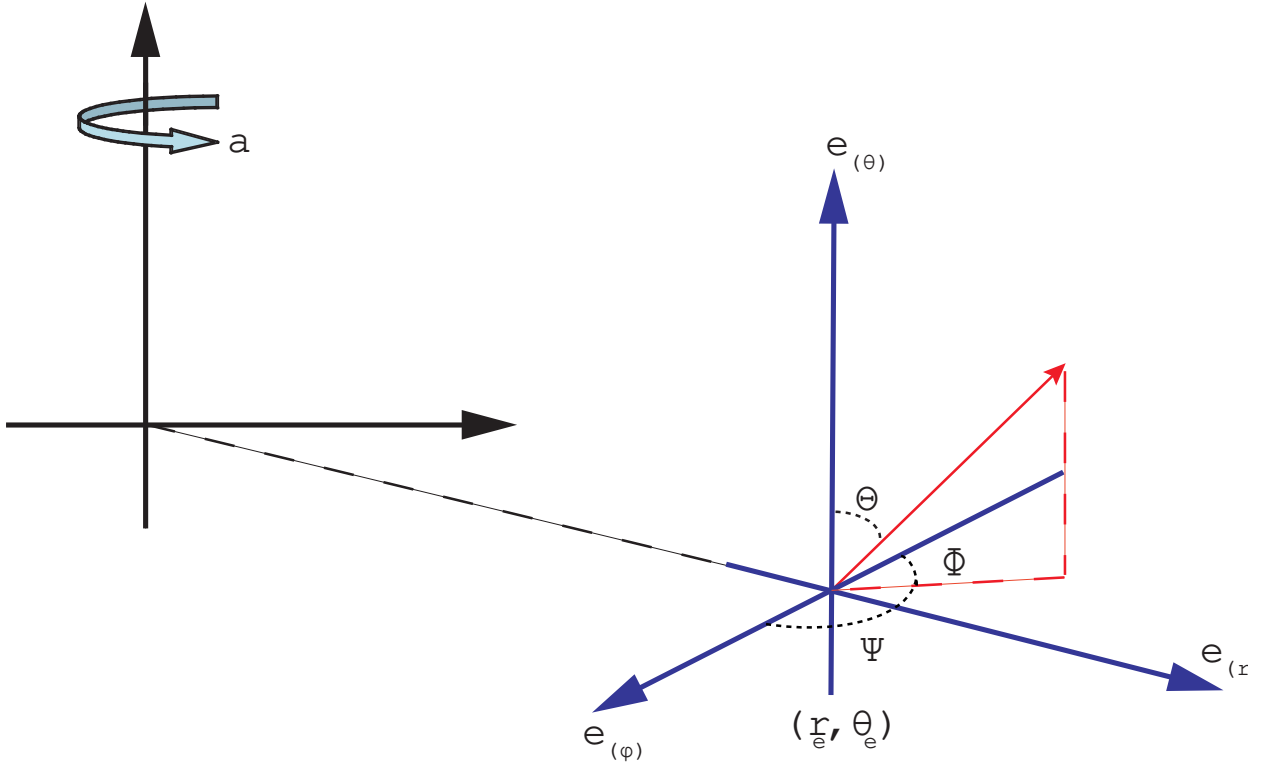}
  \end{center}
  \caption[]{The co-ordinate system associated with the fluid frame.
  The components of the photons' 4-momenta in this frame are evaluated
  through $p_{(\nu)} = e^{\mu}_{(\nu)} p_{\mu}$. From these we can
  define the angles formed by the spatial components of $p_{\mu}$ with
  the basis vectors of the fluid frame through the relations $\cos
  \Theta = p_{(\theta)} / p_{(t)}$, $\sin \Theta \sin \Phi = p_{(r)} /
  p_{(t)}$, $\sin \Theta \cos \Phi = \cos \Psi = p_{(\phi)} /
  p_{(t)}$.}
  \label{fig:3.1}
\end{figure}

Our numerical approach is as follows.  To make the problem manageable
we begin by averaging over azimuth.  Then for each grid location in $r$
and $\theta$ we generate $\sim10^{5}$ photons with a uniform
probability distribution in solid angle with respect to the fluid frame
angles $\Theta, \Phi$ (see Figure \ref{fig:3.1}).  Individual photon
orbits are defined in terms of the conserved values of angular momentum
$\lambda$ and Carter's constant $q$.  Because orbits can be traversed
in either of two senses, we also note the initial radial and polar
directions taken by the photon  with respect to the global
Boyer-Lindquist coordinates, designated $s^{e}_{r} = \pm 1$ and
$s^{e}_{\theta} = \pm 1$.  In these and in other terms, the notation
$e$ and $o$ refer to the emitted and observed coordinates of the
photon.  We must relate these quantities to the fluid frame angles for
each photon by projecting the covariant photon momenta, $p_{\mu}$ onto
a set of basis vectors,  $\mathbf{e_{(\nu)}}$ describing the rest frame
of the gas (the fluid frame, see Figure \ref{fig:3.1} and Appendix A).
This projection, $p_{(\nu)} = e^{\mu}_{(\nu)} p_{\mu}$ is related to
the fluid frame angles $\Theta, \Phi, \Psi$ via (see Figure
\ref{fig:3.1}):
\begin{equation}
  \label{eqn:3.1}
  \cos \Theta = \frac{p_{(\theta)} ( \lambda, q ; s^{e}_{r,\theta}
  )}{p_{(t)} ( \lambda, q ; s^{e}_{r,\theta} )}; \;\; \sin \Theta \sin
  \Phi = \frac{p_{(r)} ( \lambda, q ; s^{e}_{r,\theta} )}{p_{(t)} (
  \lambda, q ; s^{e}_{r,\theta} )}; \;\; \cos \Psi = - \frac{p_{(\phi)}
( \lambda, q ; s^{e}_{r,\theta} )}{p_{(t)}( \lambda, q ;
s^{e}_{r,\theta} )}; \;\;
\end{equation}

Here, for example, $p_{(\theta)} = e^{\mu}_{(\theta)} p_{\mu}$.
Unfortunately, analytic inversion of the relationships between
directions in the fluid frame, the conserved quantities $\lambda, q$
and the initial directions $s^{e}_{r,\theta}$ is not possible.
Instead, we apply a numerical algorithm described by \cite{Powell:1970}
as implemented in the NAG FORTRAN Library (Mark 21), which enables the
fast solution of the above set of equations. 
Individual photon trajectories are then determined by the analytic
integration of the null geodesic equations
\cite[see][]{Beckwith:2004,Beckwith:2005}.

Some discussion of our treatment of photons that make multiple orbits
of the black hole is necessary.  We define a ``multiple-crossing orbit"
as one that crosses the equatorial plane at least twice.  Here
we define ``orbit" as being the entire curve defined by $q$ and $\lambda$,
for either sense of traversal.  Thus, our ``multiple-crossing" category
includes what others have called ``returning radiation".   In addition,
it includes photons that originate within the plunging region, many of
which first cross the equatorial plane in the vicinity of the photon
orbit \cite[see e.g.][]{Chandrasekhar:1983}.  It is important to
distinguish those photons on multiple-crossing orbits from those that
reach infinity directly because it is likely that our assumption of
zero opacity is not valid for them.  Some, particularly those that
pass through the equatorial plane at larger radius (the ``returning
radiation" class) will be absorbed by the flow.  Others, particularly
those crossing the plane in the plunging region, even if not absorbed,
may be scattered onto capture orbits.  Unfortunately, the specific
fractions subject to these processes depend on many details of the
accretion flow.  For example, we would need to know answers to questions
such as: Is the flow hot and low-density
and capable of Compton up-scattering, or is it cool, higher density and more
absorptive?  We therefore do nothing more than count up how much energy
arriving at infinity might have been subject to effects like these.

Those photons that reach infinity are collected and used to determine
the photon transfer function, $T(r_{e}, \theta_{e}, \mu; \theta_{o},
g)$, where $\mu = \cos \Theta$, $\theta_{o}$ is the polar angle of
the trajectory at infinity, and $g$ is the photon
redshift, $g = E_{o} / E_{e} = 1 / p_{(t)}$, where $p_{(t)} =
e^{\nu}_{(t)} p_{\nu}$.  In future work we can, in principle, derive
detailed observed spectra through use of the resolved transfer
functions.  Here we aim for more modest goals by integrating over
the transfer function to determine the radiation profile (luminosity per
radial coordinate distance) in a far-away observer's rest frame:
\begin{equation}
  \label{eqn:obsprofile}
   D_{o} ( r ) \propto \int^{g_{\max}}_{g_{\min}} dg
   \int^{\theta^{\max}_{o}}_{\theta^{\min}_{o}} \sin \theta_{o} d\theta_{o}
   \int d\mu \int_{bound} {g^{2} 
   T(r_{e}, \theta_{e}, \mu; \theta_{o}, g) ||J||^{2} \alpha \sqrt
   {\gamma} \Delta r d\theta d\phi}.
 \end{equation}
That is, $D_{o} ( r )$ describes the radial profile of the radiated
energy that reaches a distant observer, integrated over the observer's
inclination angle.  To evaluate this quantity in a manner consistent
with our ansatz for the energy flux, we define the rate of energy release
in a grid-cell (measured in the fluid frame) as
$\propto ||J||^2 \alpha\sqrt{\gamma}\Delta{r} \Delta\theta \Delta\phi$
with the same proportionality constant
used to fit $D(r)$.  We further suppose that the photons are radiated
isotropically in the fluid frame.  Thus, the relation between $D_o(r)$
and $D(r)$ involves an allowance for both the fraction of photons
captured by the black hole and the Doppler shifts arising from the
transformation from the fluid frame to the coordinate frame.

\section{Photon Capture and Multiple-Crossing Fractions}
\label{sec:distrib}

\begin{figure}
  \begin{center} \leavevmode
  \includegraphics[width=0.49\textwidth]{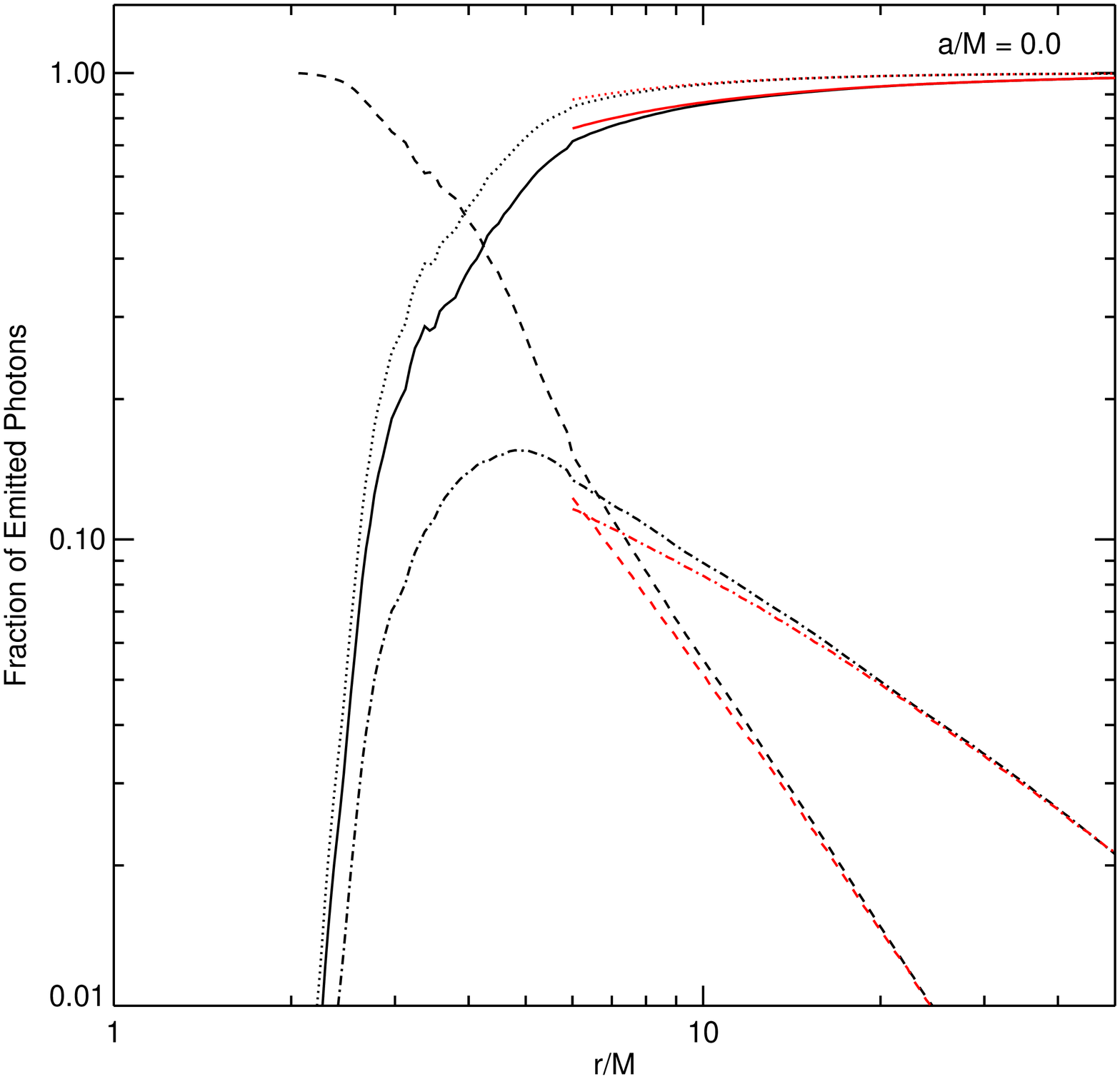}
  \includegraphics[width=0.49\textwidth]{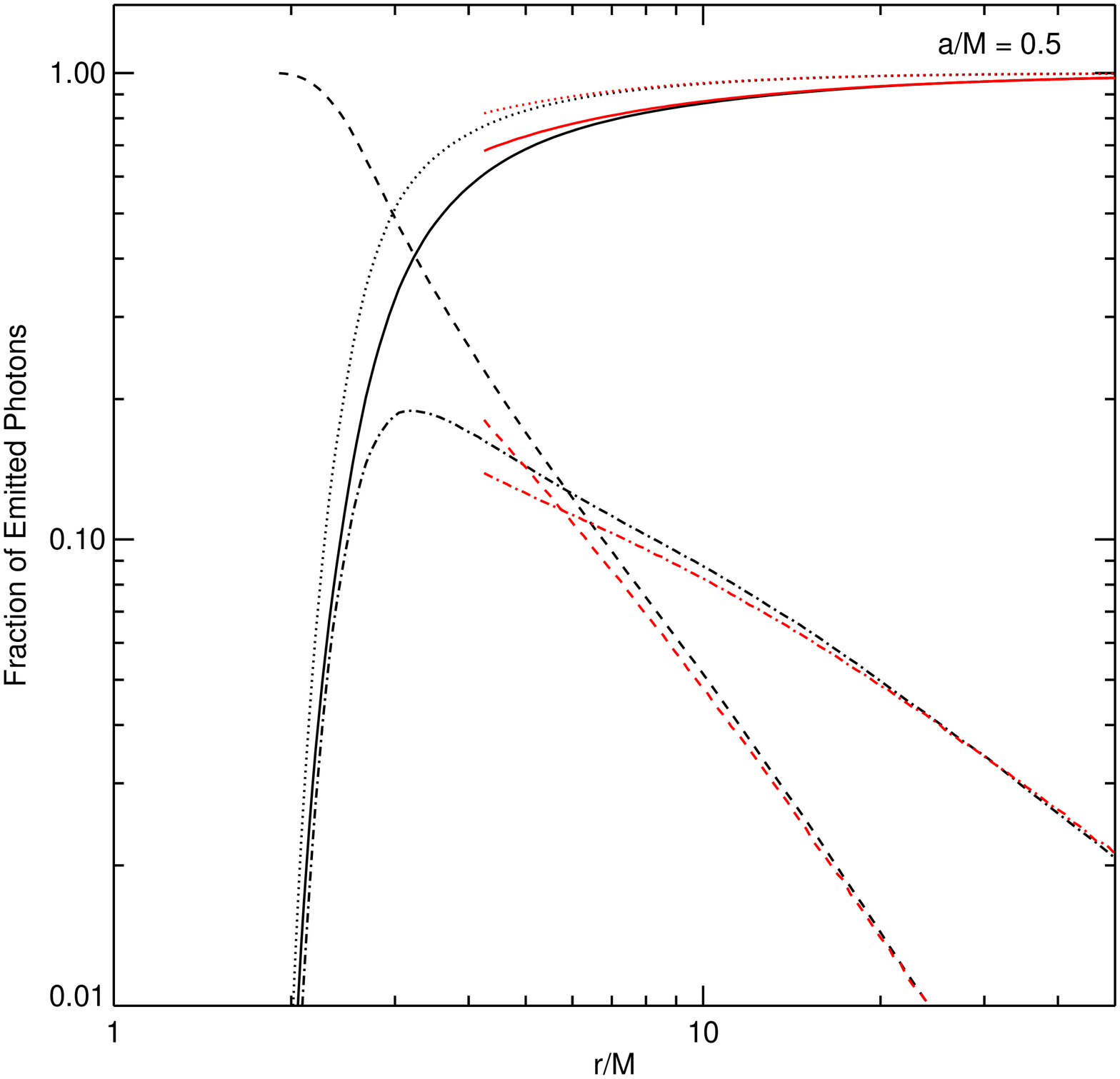}
  \includegraphics[width=0.49\textwidth]{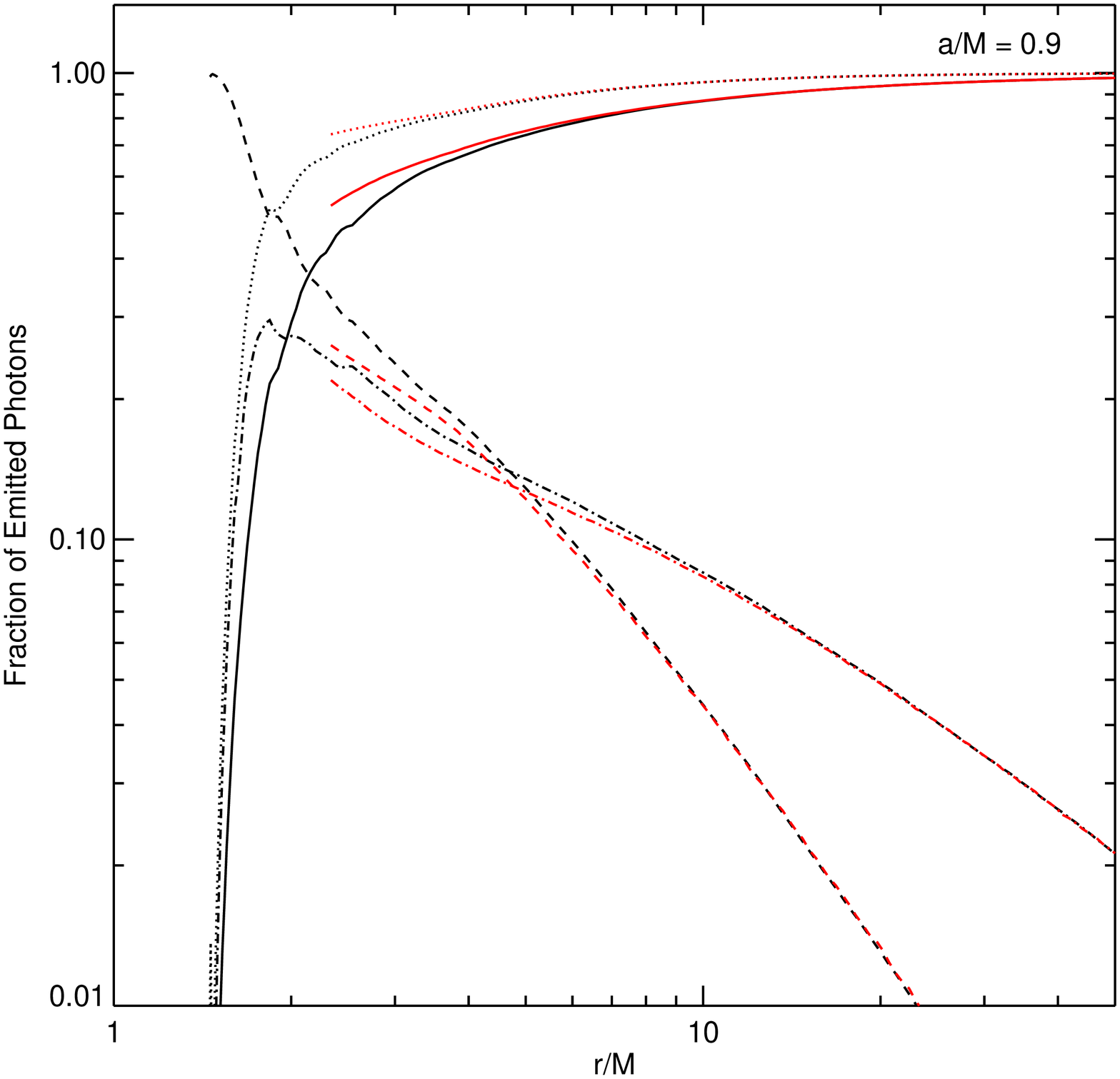}
  \includegraphics[width=0.49\textwidth]{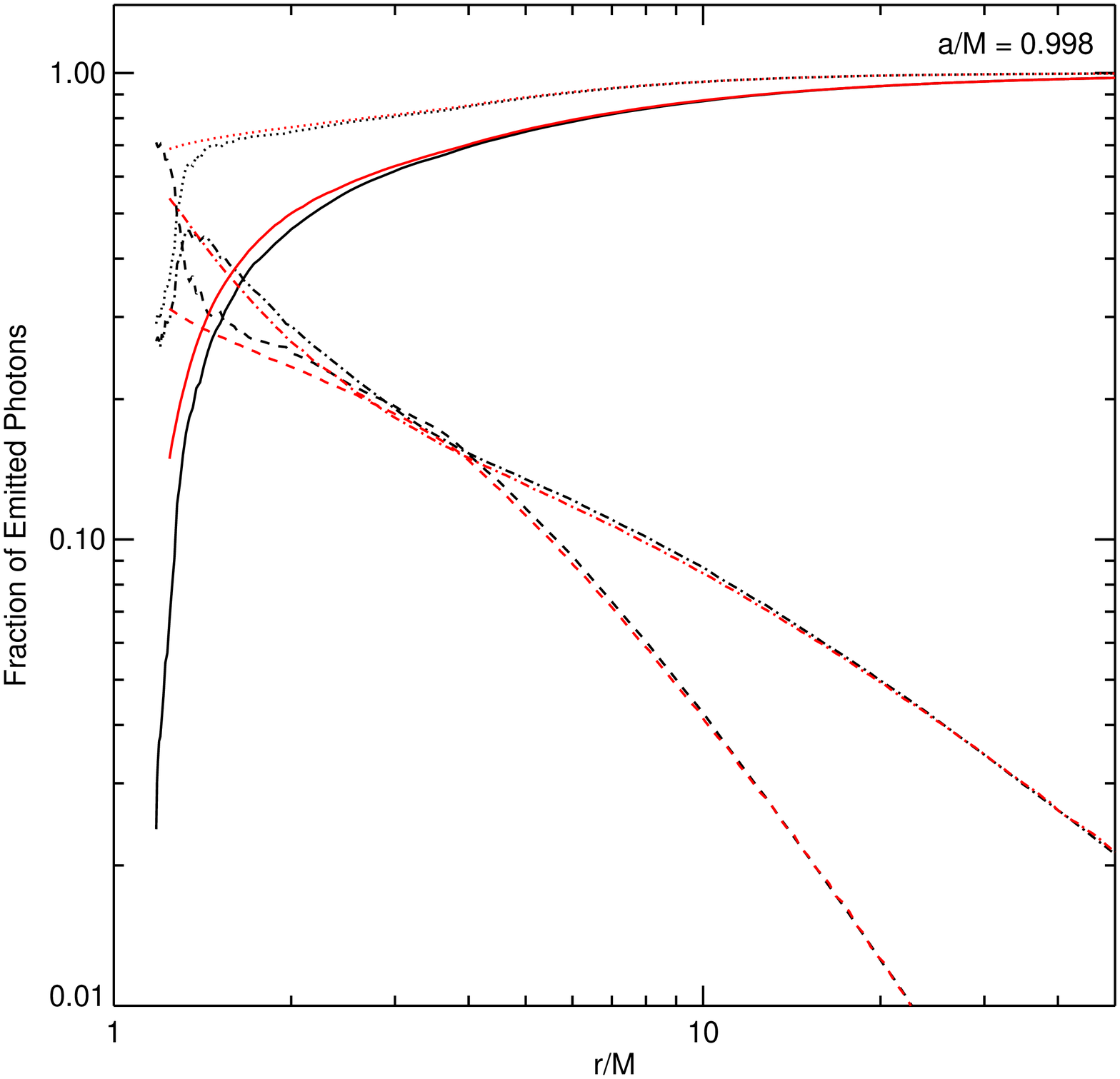}
  \end{center} \caption[]{Fractions of photons that are either captured
  by the hole or escape to infinity. The
  panels show results for $a=0$ (KD0), $a=0.5$ (KDI), $a=0.9$ (KDP) and
  $a=0.998$ (KDE). On each panel, we show: (i) the fraction of photons
  captured by the hole (dashed lines); (ii) the fraction of photons
  escaping to infinity on direct trajectories (solid lines); (iii) the
  fraction of photons escaping to infinity on multiple-crossing
  trajectories (dot-dash lines); (iv) the total fraction of photons
  escaping to infinity (dotted lines). Each panel shows these functions
  for disk-like material within the KD simulations (black lines) and a
  geometrically thin, Keplerian accretion disk extending down to
  $r_{ms}$ (red lines).} \label{fig:escapefrac} \end{figure}

With the apparatus described in \S\ref{sec:transform} we can now
answer the question: what fraction of
photons originating at a given radius are captured by the black hole, and
what fraction escape to infinity, either directly or through
multiple-crossing orbits?  Figure \ref{fig:escapefrac} displays the results
for each of the KD simulations.   The solid line corresponds to photons
that escape to infinity by direct trajectories, and the dot-dashed line
is the fraction that escapes along multiple-crossing orbits; their
sum is the dotted
line.  The dashed line is the fraction of photons that is captured by
the hole.  For comparison, we also show the results generated for a
standard geometrically thin, Keplerian accretion disk; these lines
truncate at $r_{ms}$.  Although the emission conditions at any given
$r > r_{ms}$ in the simulations are similar to those in the standard model,
they are slightly different: in the simulations, radiation is not
restricted to the equatorial plane, and the fluid velocity is not
identical to the local circular velocity.

Outside $r_{ms}(a)$ the differences between the KD models and the
standard are relatively small.  As one nears $r_{ms}$, the fraction of
photons that escape to infinity is slightly lower in the KD models;
more photons are found on multiple-crossing orbits as well.  The near perfect
agreement with the standard model beyond $r \sim 10M$ suggests that the
finite thickness of the main disk body in the simulations makes very
little difference to the probability of photon fates; in this sense the
disks are ``geometrically thin".  The only difference in the capture
rate for $r > r_{ms}$ is due to the small departures between the
fluid velocity and the circular orbital velocity.   Inside
the marginally stable orbit, of course, the standard model has no
emission.  In this region, the capture fraction rises
rapidly with decreasing radius.  In fact, the majority of emitted
photons are captured by the hole below $4M$ (KD0), $3M$ (KDI) and $2M$
(KDP).  For these simulations, the fraction of photons that travel along
multiple-crossing orbits of the hole before reaching infinity also increases
with decreasing radius, reaching a maximum of $\sim15\%$ (KD0),
$\sim30\%$ (KDI), $\sim40\%$ (KDP) at $\sim4.5, 3$ and $2.5M$,
respectively.

The results for the most rapidly rotating hole considered ($a=0.998$,
KDE) are rather different.  Again, the match with the geometrically
thin, Keplerian disk is extremely close down to $r_{ms}$. However, the
fraction of photons escaping on multiple-crossing trajectories increases
far more rapidly with decreasing radius than for the slowly rotating
black holes (for both the simulations and the standard disk model). The
majority of photons follow these trajectories for $r \lesssim 2M$,
peaking at $\sim60\%$ of the emitted photons at $\sim1.5M$. The
fraction of captured photons again increases rapidly below $r_{ms}$;
however, for the KDE simulation, this increase does not occur until
rather close to the inner boundary.

\begin{figure}
  \begin{center}
  \leavevmode
  \includegraphics[width=0.49\textwidth]{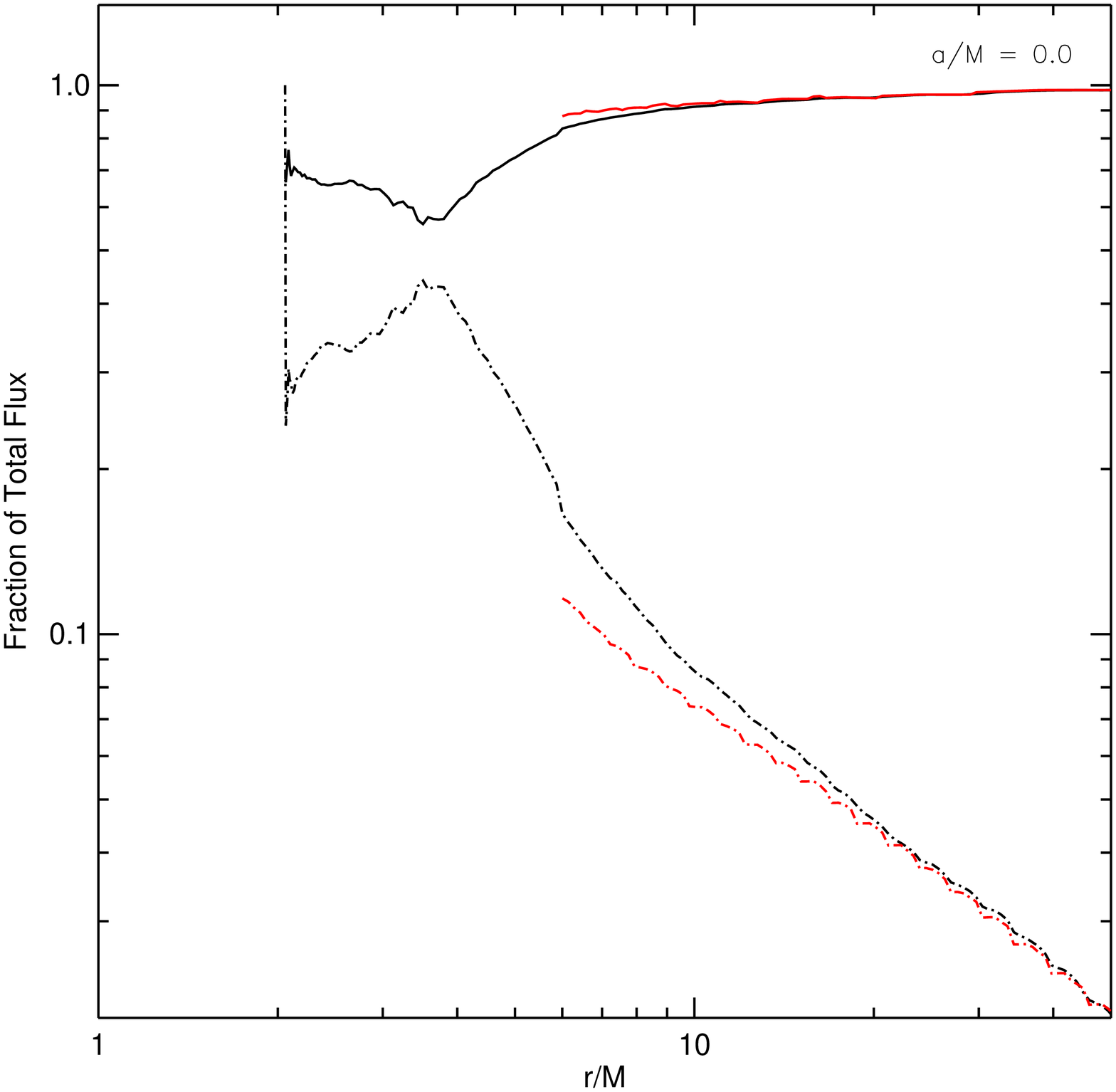}
  \includegraphics[width=0.49\textwidth]{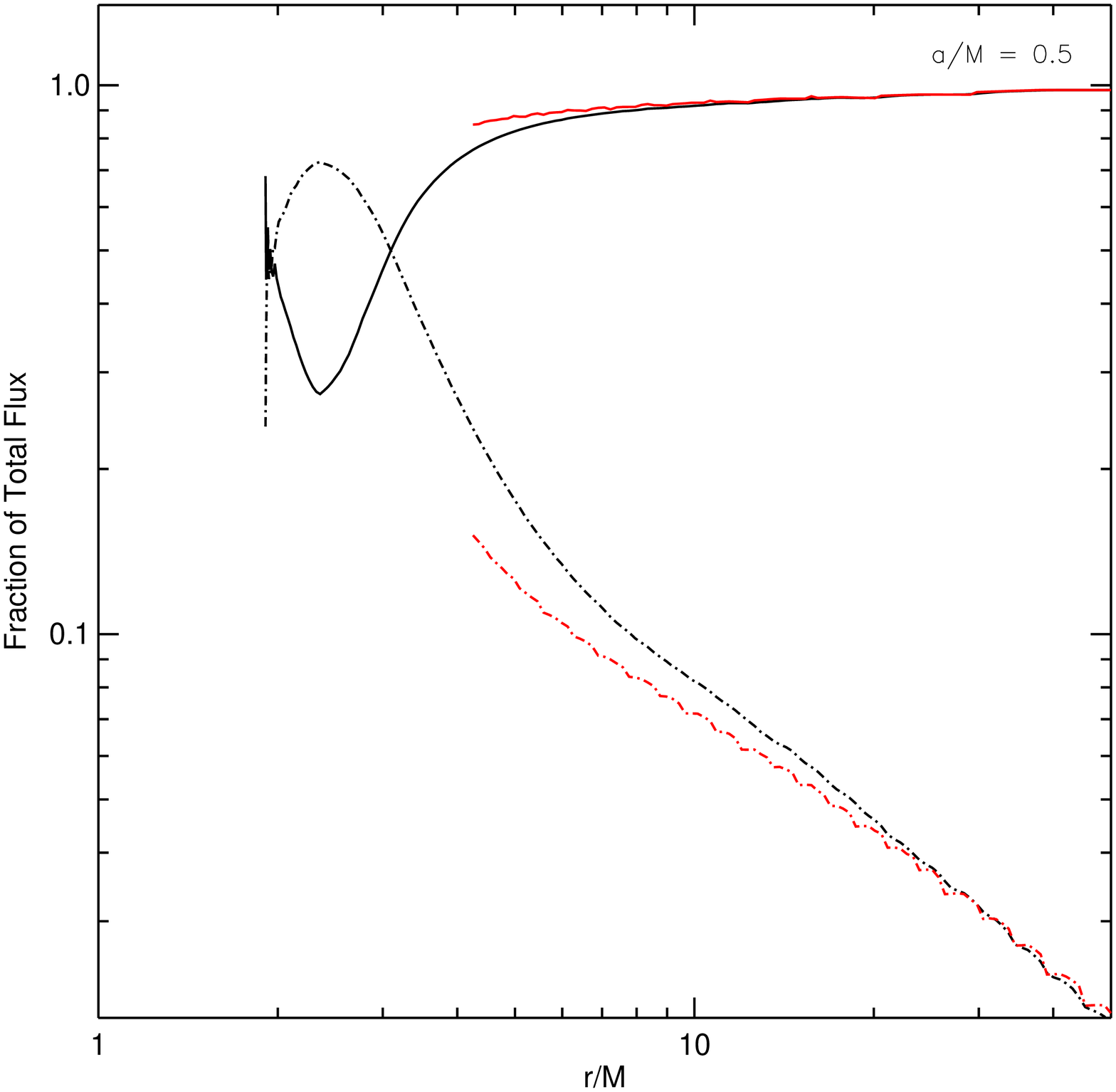}
  \includegraphics[width=0.49\textwidth]{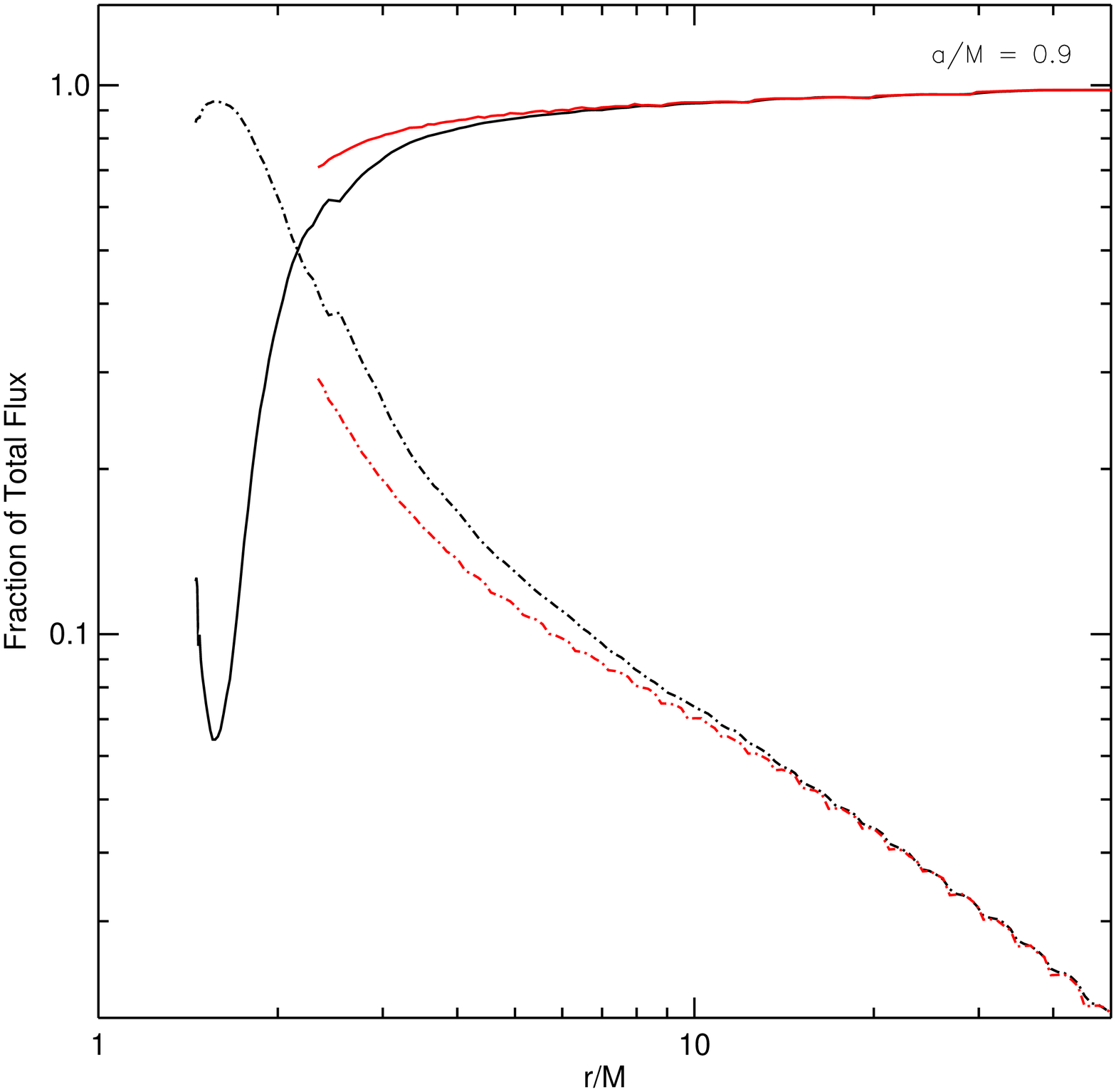}
  \includegraphics[width=0.49\textwidth]{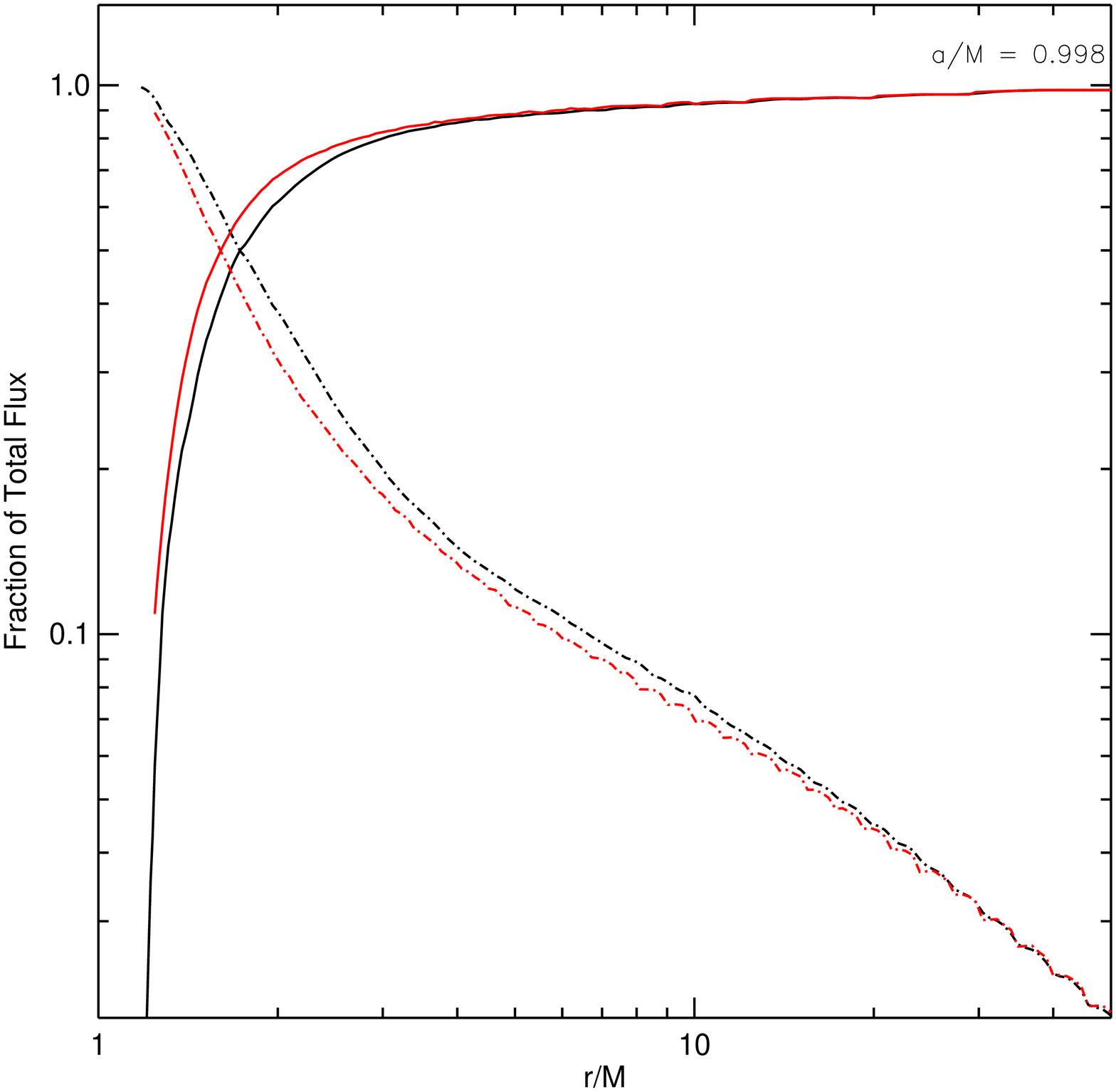}
  \end{center}
  \caption[]{Fraction of the observed flux carried to infinity by photons
that travel either directly or via multiple-crossing orbits. The
  panels show results for $a=0$ (KD0), $a=0.5$ (KDI), $a=0.9$ (KDP) and
  $a=0.998$ (KDE). Solid lines show the fraction of the observed
  flux carried by photons escaping to infinity on direct
  trajectories and dot-dash lines are photons escaping to infinity on
  multiple-crossing orbits. Each panel shows these
  functions for disk-like material within the KD simulations (black
  lines) and a geometrically thin, Keplerian accretion disk extending
  down to $r_{ms}$ (red lines).  Photons that cross the equatorial plane
  multiple times carry a
  significant fraction of the observed flux from the inner regions of
  the accretion flow.  For the accretion flows associated with the two
  rapidly rotating holes, the flux carried by multiple-crossing photons
  becomes greater than that carried by directly-escaping photons as one
  approaches the horizon.}
  \label{fig:multcrossfrac} 
\end{figure}

The plots shown in Figure \ref{fig:escapefrac} describe the probability
that an individual emitted photon will be captured by the hole or
escape to infinity (and if the latter, what type of trajectory it may
take).  These functions are not what is measured by a distant observer,
however.  Instead, the observer measures the \textit{flux} of photons
crossing his or her detector.   This is obtained from an integral over
the photon transfer function as in equation \ref{eqn:obsprofile}.  In
Figure \ref{fig:multcrossfrac}, we compare the amount of flux carried
by the direct and multiple-crossing orbit photons by plotting the
fractional amount of each as measured by such an observer. Again, there
is a close correspondence between the properties of the geometrically
thin, Keplerian accretion disk and the main disk body in the KD
simulations.  For the Schwarzschild hole the majority of the escaping
photons remain on direct orbits as one approaches the horizon.  On the
other hand, for the spinning holes, the emergent flux from
the inner regions of the accretion flow is \textit{dominated} by
contributions from photons on multiple-crossing orbits.  We caution
that a quantitative analysis of the actual fate of these photons is
likely to be highly-model dependent (see \S\ref{sec:transform}) and is
beyond the scope of our current discussion.

 \begin{figure}
  \begin{center} \leavevmode
  \includegraphics[width=0.49\textwidth]{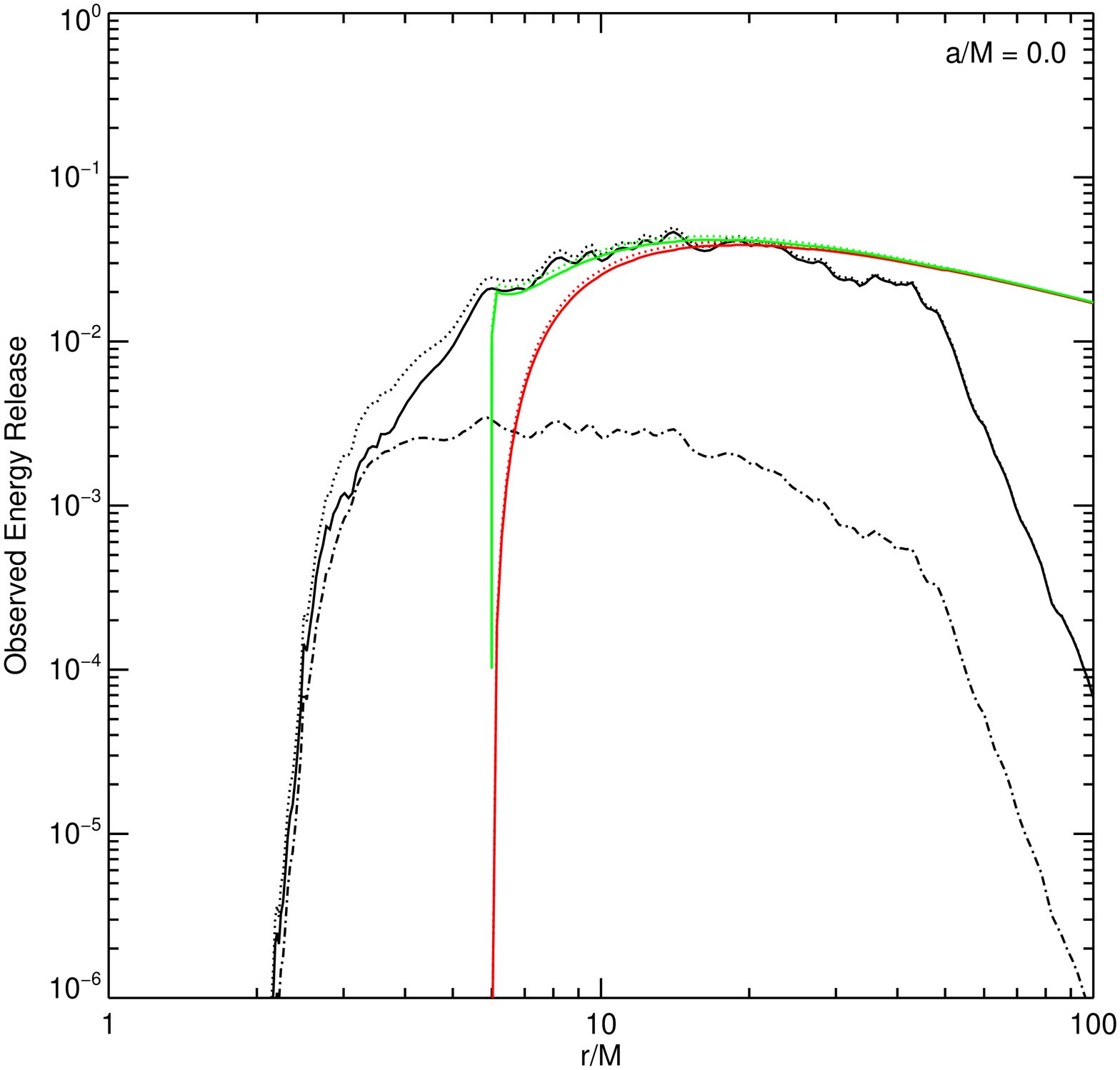}
  \includegraphics[width=0.49\textwidth]{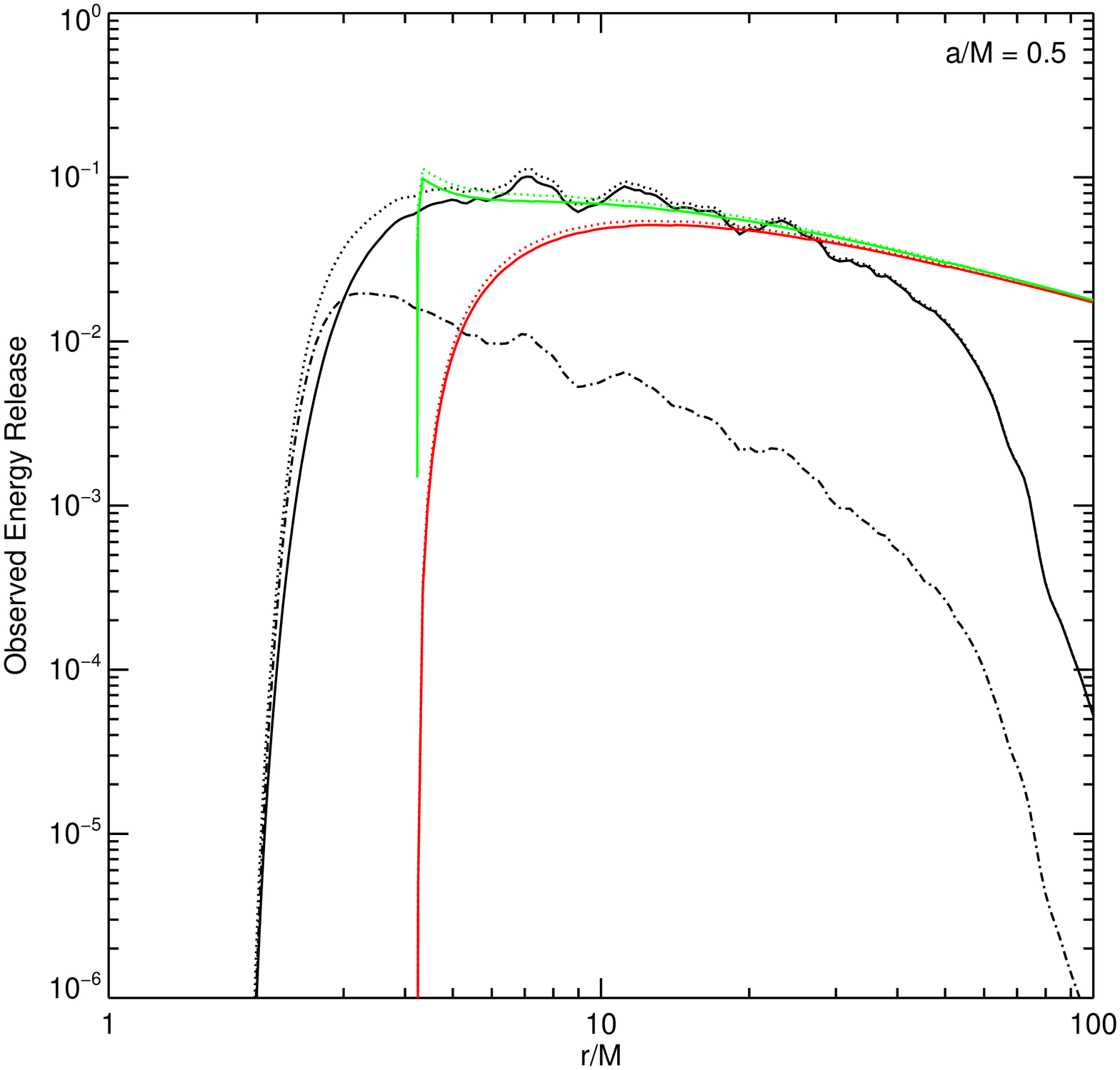}
  \includegraphics[width=0.49\textwidth]{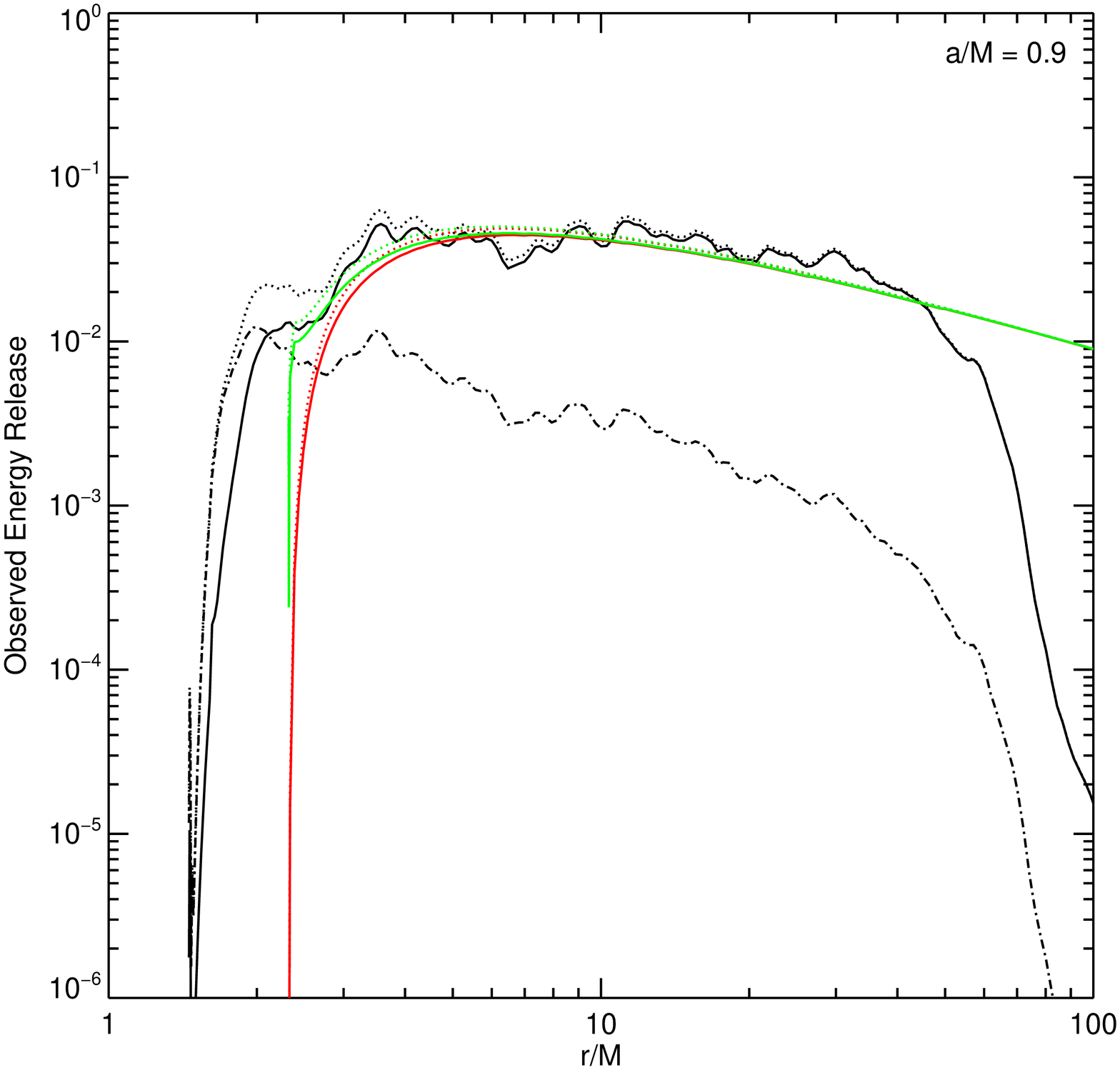}
  \includegraphics[width=0.49\textwidth]{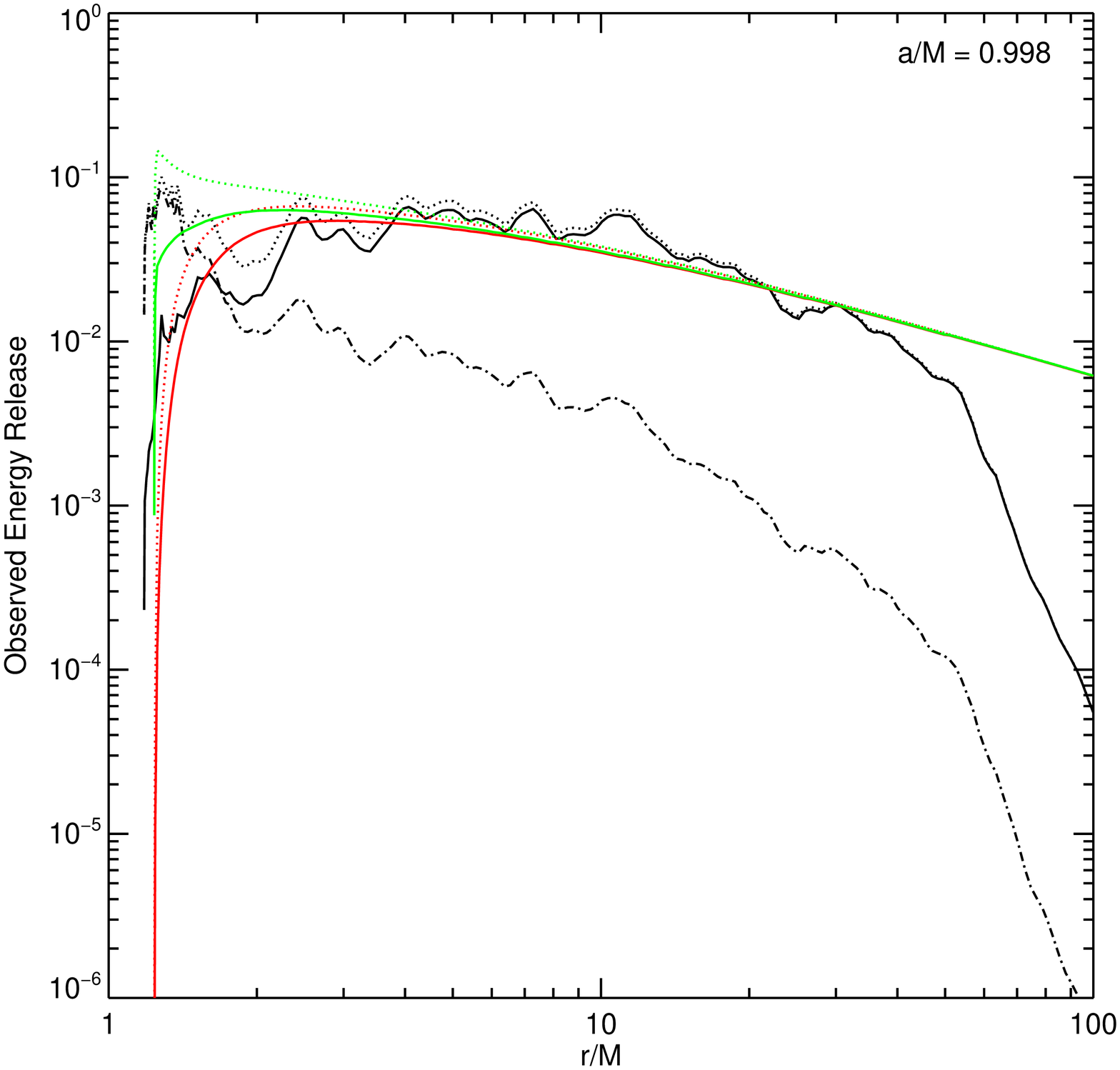}
  \end{center} \caption[]{Radiation profiles $D_o(r)$ as measured in
  the reference frame of a distant observer. Clockwise from top left,
  the panels show results for $a=0$ (KD0), $a=0.5$ (KDI), $a=0.9$ (KDP)
  and $a=0.998$ (KDE). On each panel, we show the contribution from:
  (i) photons escaping to infinity on direct trajectories (solid
  lines); (ii) photons escaping to infinity on multiple-crossing orbits
  (dot-dash lines); (iii) the total observed dissipation
  profile (dotted lines). For comparison, we also show the dissipation
  profile associated with direct photons as predicted by the standard
  relativistic disk model (red lines).} \label{fig:radnprofiles} \end{figure}

\section{Effective Radiative Efficiencies}
\label{sec:effective}

The results presented in \S\ref{sec:dissipation} showed that the dissipation rate
derived from $||J||^{2}$ predicted intrinsic radiative efficiencies that were
enhanced in comparison to those derived from the standard disk model. The majority
of this enhancement originates in the disk body just outside the marginally stable
orbit, with the plunging region contributing between $13\% - 27\%$ of the total disk
luminosity. In this section, we combine the results of \S\ref{sec:dissipation} with
the photon capture and multiple-crossing fractions discussed in \S\ref{sec:distrib}
to determine precisely how much of this enhanced dissipation arrives at the
observer, i.e., the effective radiative efficiency of the flow.

The radiation profile measured by a distant observer,
$D_{o}(r)$ is given by equation~(\ref{eqn:obsprofile}) and
is shown in Figure~\ref{fig:radnprofiles}.  This figure combines the
information on photon escape and capture (seen in 
Figures \ref{fig:escapefrac} and \ref{fig:multcrossfrac}) with the 
dissipation model seen in 
Figure \ref{fig:candchecks}. The most noticeable feature of the
profiles is the ``extra" contribution due to dissipation in the region
near and inside $r_{ms}$. The energy reaching infinity is quantified in
Tables \ref{tab:directrad} and \ref{tab:totrad}, which list the
fractional contribution to the observed flux from the same regions
defined for Table \ref{tab:currentdiss}.  Table
\ref{tab:directrad} computes this quantity for the direct photons only,
while Table \ref{tab:totrad} also includes the contributions from
photons escaping after multiple plane-crossings.  We remind the reader
that some fraction of the multiple-crossing photons radiated by a real
disk is likely to be absorbed {\it en route}, but the size of that
fraction is highly model-dependent.  As for the dissipation profile, we
define the total energy observed at infinity that originates between
radii $r_1$ and $r_2$ as
\begin{equation} 
L_{o} (r_{1},r_{2}) = \sum^{r_{2}}_{r_{1}} D_{o} (r).
\label{eqn:5.1}
\end{equation} 
For comparison we also compute the equivalent function for the standard
disk model, $L_{o}^{SM}$, using $Q$ and a set of transfer functions, $T_{SM}$,
generated using the geometric and dynamical properties of the standard
disk model.

Comparing Table~\ref{tab:directrad} and Table~\ref{tab:currentdiss}
shows that the flux delivered to the observer from the plunging region
is substantially smaller than that inferred from the
dissipation, a diminution largely attributable to the effect of photon
capture by the black hole (to zeroth order at least).  Even for the
most favorable case, the fractional contribution from the plunging
region is reduced nearly threefold in the tally of direct photons. This
picture changes quantitatively, but not qualitatively, when we include
the contribution of the multiple-orbit photons
(Table~\ref{tab:totrad}): the plunging region contributes more, but
still only a small part of the total.

As with the intrinsic efficiencies presented in Table
\ref{tab:absdiss}, the majority of the enhancement in the luminosity of
the flow over the standard disk model still originates in the
region just outside the marginally stable orbit.
Tables~\ref{tab:directrad} and \ref{tab:totrad} show that the
enhancement in this region over the standard disk model is
again maximized for the intermediate spin case, and that the level of
enhancement is somewhat reduced in comparison to the intrinsic
efficiencies.

That the increased efficiency is greatest for intermediate spin is the
result of a trade-off between dissipation and capture, both of which
increase with increasing black hole spin.  At slow spin, there is
relatively little additional dissipation, while at very high spin, so
much of the light from the inner part of the accretion flow is captured
that these regions contribute little to what can be seen at infinity.
Put another way, at high spin, the portion of the disk affected most
strongly by elimination of the zero-stress boundary condition is so
deep in the gravitational well of the black hole that it is only barely
visible from the outside.

For both of the two low-spin cases, we can quote reasonably
well-defined predictions for the simulations.  Both are clearly in
inflow equilibrium \cite[]{Krolik:2005} and, as
Figure~\ref{fig:radnprofiles} shows, the radial fluctuation levels are
low enough that radial integrations for these snapshots likely give a
fair representation of the time-averaged condition.  The increase in
the effective radiation efficiency relative
to the standard model is $\Delta \epsilon = \epsilon_{||J||^{2}} -
\epsilon_{SM} \simeq 0.012$--0.014 (a fractional increase of $22\%$--$25\%$) for
the non-rotating black hole, where the -- denotes the range spanned between
the direct photon only and total photon numbers.  The increase is considerably
greater for $a = 0.5$, $\Delta \epsilon \simeq 0.050$--0.063 ($65\%$--$80\%$).
Only a small minority
of this light comes from within the plunging region; most comes from
the disk body at $r > r_{ms}$.

    The situation changes in several respects when the black hole rotates more
rapidly.   As already remarked, substantially more of of the energy released
within the flow is captured by the black hole than in the slowly rotating
case.  Because multiple-crossing photons also play a bigger role, there
is greater uncertainty in the efficiency due to our uncertainty in how
much of the energy they carry reaches infinity.  There are also two ways
in which the numerical situation is different.  As
shown in \cite{Krolik:2005}, the KDE model is not in inflow equilibrium,
so interpretation of this radial emission profile as a sample of the
time-average is a bit dubious.  In addition, even in KDP, which is in inflow
equilibrium, the radial fluctuations are so large that the uncertainty
in using these data alone as an estimate of the time-average must be quite
sizable; factors of several might be expected.   Nominally, we find that,
when $a/M=0.9$, the efficiency might be increased in absolute terms by
0.025--0.037 ($18\%$--$26\%$ fractionally), and when $a/M = 0.998$,
the increase could be 0.04--0.12 in absolute terms ($16\%$--$42\%$).
We stress, however, that for all these reasons, the
high-spin efficiency estimates are considerably more uncertain than the
low-spin numbers.

\begin{deluxetable}{lrcccccc}
\tablecolumns{5}
\tablewidth{0pc}
\tablecaption{Observed Luminosities (Direct Photons Only)}
\tablehead{\colhead{Model}          &
           \colhead{$a/M$}         &
           \colhead{$\frac{L_{o}(r_{in},r_{ms})}{L_{o}(r_{in},\infty)}$}     &
           \colhead{$\frac{L_{o}(r_{ms},\infty)}{L_{o}(r_{in},\infty)}$} &
           \colhead{$\frac{L_{o}(r_{in},\infty)}{L_{o}^{SM}(r_{ms},\infty)}$}
}
\startdata
KD0 &    0.0 &     0.05 &      0.95 &       1.22 \\
KDI &      0.5 &     0.09 &      0.91 &       1.66 \\
KDP &      0.9 &     0.02 &      0.98 &       1.18 \\
KDE &      0.998 &  $<$0.01 &  $>$ 0.99 &       1.16 \\
\enddata
\label{tab:directrad}
\end{deluxetable}

\begin{deluxetable}{lrcccccc}
\tablecolumns{5}
\tablewidth{0pc}
\tablecaption{Observed Luminosities (All Photons)}
\tablehead{\colhead{Model}          &
           \colhead{$a/M$}         &
           \colhead{$\frac{L_{o}(r_{in},r_{ms})}{L_{o}(r_{in},\infty)}$}     &
           \colhead{$\frac{L_{o}(r_{ms},\infty)}{L_{o}(r_{in},\infty)}$} &
            \colhead{$\frac{L_{o}(r_{in},\infty)}{L_{o}^{SM}(r_{ms},\infty)}$}
}
\startdata
KD0 &    0.0 &     0.06 &      0.94 &       1.25 \\
KDI &      0.5 &      0.13 &      0.87 &       1.80 \\
KDP &     0.9 &     0.05 &      0.95 &       1.25 \\
KDE &      0.998 &     0.06 &      0.94 &       1.41 \\
\enddata
\label{tab:totrad}
\end{deluxetable}

\begin{deluxetable}{lrcccccc}
\tablecolumns{5}
\tablewidth{0pc}
\tablecaption{Observed Radiative Efficiencies (Direct Photons)}
\tablehead{\colhead{Model}          &
            \colhead{$a/M$}         &
            \colhead{$\epsilon_{SM} (r_{ms},\infty)$}         &
            \colhead{$\epsilon_{SSM} (r_{ms},\infty)$}         &
            \colhead{$\epsilon_{||J||^{2}} (r_{in},\infty)$}         &
}
\startdata
KD0 &    0.0 &     0.055  &  0.062 &     0.067 \\
KDI &      0.5 &     0.077  &   0.111 &      0.127 \\
KDP &      0.9 &      0.137  &   0.143 &      0.162 \\
KDE &      0.998 &      0.250  &   0.296 &      0.291 \\
\enddata
\label{tab:directobseff}
\end{deluxetable}

\begin{deluxetable}{lrcccccc}
\tablecolumns{5}
\tablewidth{0pc}
\tablecaption{Observed Radiative Efficiencies (All Photons)}
\tablehead{\colhead{Model}          &
            \colhead{$a/M$}         &
            \colhead{$\epsilon_{SM} (r_{ms},\infty)$}         &
            \colhead{$\epsilon_{SSM} (r_{ms},\infty)$}         &
            \colhead{$\epsilon_{||J||^{2}} (r_{in},\infty)$}         &
}
\startdata
KD0 &    0.0 &     0.056  &  0.063 &     0.070 \\
KDI &      0.5 &     0.079  &   0.116 &      0.142 \\
KDP &      0.9 &      0.145  &   0.153 &      0.182 \\
KDE &      0.998 &      0.290  &   0.389 &      0.411 \\
\enddata
\label{tab:totobseff}
\end{deluxetable}

\section{Summary and Discussion}
\label{sec:summ}

Simulations of accretion flows in which the internal torques arise
self-consistently from the underlying physics can, in principle, lead
to detailed predictions for observations of black hole systems.  A
potentially observable effect that arises immediately in such
simulations is the presence of magnetic stress both across the
marginally stable orbit and into the plunging region of the accretion
flow (Krolik et al. 2005).  This behavior stands in marked contrast to the
assumptions of the standard accretion disk model and raises the
prospect of dissipation occurring within the plunging region.   Where
there is dissipation, there may well be radiation.  The present
simulations, while capturing the overall dynamics of the flows, do not
model dissipation.  To begin to evaluate the observational
consequences we must, therefore adopt a prescription for dissipation.
Motivated by the suggestions of \cite{Rosner:1978}, \cite{Machida:2003},
\cite{Hirose:2004}, and \cite{Hirose:2005}, along with a
naive expectation from Ohm's Law, we assume a simple relationship between
dissipation within the accretion flow and the magnetic $4$-current
density, $||J||^{2}$.  Comparing the radial profile of the
shell-integrated $||J||^{2}$ with the dissipation function $Q$ demanded by
energy conservation within the standard model reveals a good match
between these two quantities for $r_{ms} < r < 20M$; an even better
match is made with the ``stressed standard model" (Agol \& Krolik 2000),
an adjustment of the standard model that allows for non-zero stresses at
$r=r_{ms}$.

Enhanced dissipation is one thing; how much of the energy radiated as a
consequence of this dissipation reaches distant observers is another.
To begin to answer this question we have adopted the simplest possible
approximation to the radiation transfer problem: instantaneous conversion
of heat to radiation and zero opacity everywhere.  Photon trajectories
can then be computed and a photon transfer function derived to determine
which photons reach infinity, and with what Doppler shifts.  In the
end, we find that the luminosity produced per unit rest-mass
accreted, that is the effective radiative efficiency, can be enhanced by
anywhere from tens of percent to a factor of (nearly) two, with the
maximum occurring at intermediate $a/M$. Although our numbers are much less
well-defined for the two cases of rapid
rotation we studied, we find that their effective radiative
efficiency is enhanced (in proportionate terms) less than
for black holes of intermediate spin because
so much of the additional dissipation takes place very deep in the
black hole's gravitational potential: much of the energy released is
captured by the black hole.  At any spin,
the majority of the additional light comes from the region a short
way outside the marginally stable orbit.

Granted these results, the radiative efficiency of accreting black
holes may be considerably greater than previously thought.  If so,
estimates of population-mean spin
parameters based on inferred efficiencies \cite[e.g., as for AGN by][]
{Elvis:2002}, may substantially overestimate the typical spin
of accreting black holes.

Additional luminosity from the innermost part of the accretion flow
should also alter predictions of spectra from accreting black holes.
If the energy is thermalized, the thermal peak (at $\sim 1$~keV in
Galactic black holes, $\sim 10$~eV in AGN) will be pushed to somewhat
higher energies \cite[]{Agol:2000}.   On the other hand, it is also
possible that some of the energy is released in places where the 
density and optical depth are too low to accomplish thermalization.
Strengthening of the ``coronal", i.e., hard X-ray, emission would
then be the likely consequence.  Firming up these predictions,
however, demands a more comprehensive approach to the gas's
thermodynamics and opacity than can be supported by the data of
the simulations performed to date.

\acknowledgements{ This work was supported by NSF grant
PHY-0205155 and NASA grant NNG04GK77G (JFH),
and by NSF grants AST-0205806 and AST-0313031
(JHK).  The GRMHD simulations described here were carried out on the DataStar
system at SDSC.  The photon transfer calculations were carried in part at the
University of Durham.  KB thanks C. Done, M. Gierli\'nski, S. Matt and
A.A. Constandache for useful discussions and advice.}

\section{Appendix A}

\label{sec:appa}

The calculation of the photon transfer functions requires the
introduction of a set of basis vectors describing the local rest frame
of the fluid (the ``fluid frame"). Such a tetrad set was presented by
\cite{Krolik:2005}, who used a Gram-Schmidt orthonormalization
procedure to construct it. Unfortunately, some of the expressions
given in that work were
incorrectly transcribed. The correct version is as follows:
\begin{align}
  e^{\mu}_{(t)} = U^{t} \left( 1, v^{r}, v^{t}, v^{\phi} \right) \\
  e^{\mu}_{(r)} = \frac{1}{N_{2}} \left( -g^{tt} k_{1} - \ell g^{t \phi} k_{1}, g^{rr} , 0 ,
  -g^{t \phi} k_{1} - g^{\phi \phi} \ell k_{1} \right) \\
  e^{\mu}_{(\theta)} = \frac{1}{N_{3}} \left( U^{\theta} g^{tt} +  U^{\theta} g^{t\phi} \ell ,
  g^{rr} k_{1} k_{2} U^{\theta}, -g^{\theta \theta} k_{3},
  U^{\theta} g^{t \phi} + U^{\theta} g^{\phi \phi} \ell \right) \\
  e^{\mu}_{(\phi)} = \frac{1}{N_{1}} \left( -\ell, 0, 0, 1 \right)
\end{align}
Here:
\begin{align}
  \ell = U_{\phi} / U_{t}\\
  k_{1} = U^{r} / \left( U^{t} + U^{\phi} \ell \right)\\
  k_{2} = \left( g^{tt} + 2g^{t \phi} \ell +g^{\phi \phi} \ell^{2} \right) / g^{rr} \\
  k_{3} = \left( U^{t} + k_{1} k_{2} U^{r} + \ell U^{\phi} \right)^{-1}\\
  N_{1} = \sqrt{ g_{tt} \ell^{2} - 2 g_{t \phi} \ell +g_{\phi \phi}}\\
  N_{2} = \sqrt{ g^{tt} k^{2}_{1} + 2 g^{t \phi} k^{2}_{1} \ell + g^{rr} + g^{\phi \phi} k^{2}_{1} \ell} \\
  N_{3} = \sqrt{ \left(U^{\theta}k_{3}\right)^2
               \left( g^{tt} + 2 g^{t \phi} \ell + g^{rr} k^{2}_{1} k^{2}_{2}
                            + g^{\phi \phi} \ell^{2} \right) + g^{\theta \theta}}
\end{align}


\begin{thebibliography}{24}
\expandafter\ifx\csname natexlab\endcsname\relax\def\natexlab#1{#1}\fi
\expandafter\ifx\csname href\endcsname\relax
  \def\href#1#2{}\fi
\expandafter\ifx\csname urllinklabel\endcsname\relax
  \def\urllinklabel{[LINK]}\fi
\expandafter\ifx\csname adsurllinklabel\endcsname\relax
  \def\adsurllinklabel{[ADS]}\fi

\bibitem[{{Agol} \& {Krolik}(2000)}]{Agol:2000}
{Agol}, E. \& {Krolik}, J.~H. 2000, \apj, 528, 161
 \href{http://ukads.nottingham.ac.uk/cgi-bin/nph-bib_query?bibcode=2000ApJ...5%
28..161A&db_key=AST}{\adsurllinklabel}

\bibitem[{{Balbus} \& {Hawley}(1998)}]{Balbus:1998}
{Balbus}, S.~A. \& {Hawley}, J.~F. 1998, Reviews of Modern Physics, 70, 1
 \href{http://ukads.nottingham.ac.uk/cgi-bin/nph-bib_query?bibcode=1998RvMP...%
70....1B&db_key=AST}{\adsurllinklabel}

\bibitem[{{Balbus} \& {Papaloizou}(1999)}]{Balbus:1999}
{Balbus}, S.~A. \& {Papaloizou}, J.~C.~B. 1999, \apj, 521, 650
 \href{http://adsabs.harvard.edu/cgi-bin/nph-bib_query?bibcode=1999ApJ...521..%
650B&db_key=AST}{\adsurllinklabel}

\bibitem[{{Beckwith} \& {Done}(2004)}]{Beckwith:2004}
{Beckwith}, K. \& {Done}, C. 2004, \mnras, 352, 353
 \href{http://ukads.nottingham.ac.uk/cgi-bin/nph-bib_query?bibcode=2004MNRAS.3%
52..353B&db_key=AST}{\adsurllinklabel}

\bibitem[{{Beckwith} \& {Done}(2005)}]{Beckwith:2005}
---. 2005, \mnras, 359, 1217
 \href{http://ukads.nottingham.ac.uk/cgi-bin/nph-bib_query?bibcode=2005MNRAS.3%
59.1217B&db_key=AST}{\adsurllinklabel}

\bibitem[{{Chandrasekhar}(1983)}]{Chandrasekhar:1983}
{Chandrasekhar}, S. 1983, {The mathematical theory of black holes} (Oxford:
  Oxford University Press)
 \href{http://ukads.nottingham.ac.uk/cgi-bin/nph-bib_query?bibcode=1983mtbh.bo%
ok.....C&db_key=AST}{\adsurllinklabel}

\bibitem[{{De Villiers} \& {Hawley}(2003{\natexlab{a}})}]{De-Villiers:2003}
{De Villiers}, J.-P. \& {Hawley}, J.~F. 2003{\natexlab{a}}, \apj, 589, 458
 \href{http://ukads.nottingham.ac.uk/cgi-bin/nph-bib_query?bibcode=2003ApJ...5%
89..458D&db_key=AST}{\adsurllinklabel}

\bibitem[{{De Villiers} \& {Hawley}(2003{\natexlab{b}})}]{De-Villiers:2003a}
---. 2003{\natexlab{b}}, \apj, 592, 1060
 \href{http://ukads.nottingham.ac.uk/cgi-bin/nph-bib_query?bibcode=2003ApJ...5%
92.1060D&db_key=AST}{\adsurllinklabel}

\bibitem[{{De Villiers} {et~al.}(2003){De Villiers}, {Hawley}, \&
  {Krolik}}]{De-Villiers:2003b}
{De Villiers}, J.-P., {Hawley}, J.~F., \& {Krolik}, J.~H. 2003, \apj, 599, 1238
 \href{http://ukads.nottingham.ac.uk/cgi-bin/nph-bib_query?bibcode=2003ApJ...5%
99.1238D&db_key=AST}{\adsurllinklabel}

\bibitem[{{De Villiers} {et~al.}(2005){De Villiers}, {Hawley}, {Krolik}, \&
  {Hirose}}]{De-Villiers:2005}
{De Villiers}, J.-P., {Hawley}, J.~F., {Krolik}, J.~H., \& {Hirose}, S. 2005,
  \apj, 620, 878
 \href{http://ukads.nottingham.ac.uk/cgi-bin/nph-bib_query?bibcode=2005ApJ...6%
20..878D&db_key=AST}{\adsurllinklabel}

\bibitem[{{Elvis} {et~al.}(2002){Elvis}, {Risaliti}, \&
  {Zamorani}}]{Elvis:2002}
{Elvis}, M., {Risaliti}, G., \& {Zamorani}, G. 2002, \apjl, 565, L75
 \href{http://adsabs.harvard.edu/cgi-bin/nph-bib_query?bibcode=2002ApJ...565L.%
.75E&db_key=AST}{\adsurllinklabel}

\bibitem[{{Gammie}(1999)}]{Gammie:1999}
{Gammie}, C.~F. 1999, \apjl, 522, L57
 \href{http://adsabs.harvard.edu/cgi-bin/nph-bib_query?bibcode=1999ApJ...522L.%
.57G&db_key=AST}{\adsurllinklabel}

\bibitem[{{Hirose} {et~al.}(2004){Hirose}, {Krolik}, {De Villiers}, \&
  {Hawley}}]{Hirose:2004}
{Hirose}, S., {Krolik}, J.~H., {De Villiers}, J.-P., \& {Hawley}, J.~F. 2004,
  \apj, 606, 1083
 \href{http://ukads.nottingham.ac.uk/cgi-bin/nph-bib_query?bibcode=2004ApJ...6%
06.1083H&db_key=AST}{\adsurllinklabel}

\bibitem[{{Hirose} {et~al.}(2006){Hirose}, {Krolik}, \& {Stone}}]{Hirose:2005}
{Hirose}, S., {Krolik}, J.~H., \& {Stone}, J.~M. 2006, \apj, 640, 901
 \href{http://adsabs.harvard.edu/cgi-bin/nph-bib_query?bibcode=2006ApJ...640..%
901H&db_key=AST}{\adsurllinklabel}

\bibitem[{{Krolik}(1999{\natexlab{a}})}]{Krolik:1999}
{Krolik}, J.~H. 1999{\natexlab{a}}, {Active galactic nuclei : from the central
  black hole to the galactic environment} (Princeton University Press)
 \href{http://ukads.nottingham.ac.uk/cgi-bin/nph-bib_query?bibcode=1999agnc.bo%
ok.....K&db_key=AST}{\adsurllinklabel}

\bibitem[{{Krolik}(1999{\natexlab{b}})}]{Krolik:1999a}
---. 1999{\natexlab{b}}, \apjl, 515, L73
 \href{http://ukads.nottingham.ac.uk/cgi-bin/nph-bib_query?bibcode=1999ApJ...5%
15L..73K&db_key=AST}{\adsurllinklabel}

\bibitem[{{Krolik} \& {Hawley}(2002)}]{Krolik:2002}
{Krolik}, J.~H. \& {Hawley}, J.~F. 2002, \apj, 573, 754
 \href{http://ukads.nottingham.ac.uk/cgi-bin/nph-bib_query?bibcode=2002ApJ...5%
73..754K&db_key=AST}{\adsurllinklabel}

\bibitem[{{Krolik} {et~al.}(2005){Krolik}, {Hawley}, \& {Hirose}}]{Krolik:2005}
{Krolik}, J.~H., {Hawley}, J.~F., \& {Hirose}, S. 2005, \apj, 622, 1008
 \href{http://ukads.nottingham.ac.uk/cgi-bin/nph-bib_query?bibcode=2005ApJ...6%
22.1008K&db_key=AST}{\adsurllinklabel}

\bibitem[{{Machida} \& {Matsumoto}(2003)}]{Machida:2003}
{Machida}, M. \& {Matsumoto}, R. 2003, \apj, 585, 429
 \href{http://adsabs.harvard.edu/cgi-bin/nph-bib_query?bibcode=2003ApJ...585..%
429M&db_key=AST}{\adsurllinklabel}

\bibitem[{{Novikov} \& {Thorne}(1973)}]{Novikov:1973}
{Novikov}, I.~D. \& {Thorne}, K.~S. in , {Black Holes: Les Astres Occlus}, ed.
  C.~{DeWitt}B.~{DeWitt} (New York; Gordon and Breach)


\bibitem[{{Page} \& {Thorne}(1974)}]{Page:1974}
{Page}, D.~N. \& {Thorne}, K.~S. 1974, \apj, 191, 499
 \href{http://adsabs.harvard.edu/cgi-bin/nph-bib_query?bibcode=1974ApJ...191..%
499P&db_key=AST}{\adsurllinklabel}

\bibitem[{{Powell}(1970)}]{Powell:1970}
{Powell}, M.~J.~D. 1970, in {Numerical methods for nonlinear algebraic
  equations}, ed. P.~{Rabinowitz} ({New York: Gordon \& Breach})
 \href{https://virgo.lib.virginia.edu/uhtbin/cgisirsi/dWl2DPjD36/UVA-LIB/31693%
0106/9}{\urllinklabel}

\bibitem[{{Rosner} {et~al.}(1978){Rosner}, {Golub}, {Coppi}, \&
  {Vaiana}}]{Rosner:1978}
{Rosner}, R., {Golub}, L., {Coppi}, B., \& {Vaiana}, G.~S. 1978, \apj, 222, 317
 \href{http://ukads.nottingham.ac.uk/cgi-bin/nph-bib_query?bibcode=1978ApJ...2%
22..317R&db_key=AST}{\adsurllinklabel}

\bibitem[{{Thorne}(1974)}]{Thorne:1974}
{Thorne}, K.~S. 1974, \apj, 191, 507
 \href{http://ukads.nottingham.ac.uk/cgi-bin/nph-bib_query?bibcode=1974ApJ...1%
91..507T&db_key=AST}{\adsurllinklabel}

\end{thebibliography}
\end{document}